\documentclass[aps,preprintnumbers,showpacs,nofootinbib]{revtex4}
\usepackage{subfig}
\usepackage{graphicx}
\usepackage{dcolumn}
\usepackage{epsfig}
\usepackage{bm}
\usepackage{natbib}
\newcommand{\eq}{\begin{eqnarray}}
\newcommand{\en}{\end{eqnarray}}

\newcommand{\ba}[1]{\begin{eqnarray} \label{(#1)}}
\newcommand{\ea}{\end{eqnarray}}

\newcommand{\newc}{\newcommand}
\newc{\lra}{\leftrightarrow}
\newc{\beq}{\begin{equation}}
\newc{\eeq}{\end{equation}}
\newc{\barr}{\begin{eqnarray}}
\newc{\earr}{\end{eqnarray}}
  \def\vbf{\mbox{\boldmath $\upsilon$}}
\begin{document}

\topmargin -0.50in
\title { Searching for light WIMPS via their  interaction with electrons} 
\author{J.D. Vergados}
\affiliation{ University of Ioannina, Ioannina, Gr 451 10, Greece.},
\begin{abstract}
We  consider light WIMP searches  involving the detection of recoiling electrons.
\end{abstract}


\date{\today}
\begin{abstract}
In the present work we examine the possibility of detecting electrons in dark matter searches employing  for detectors  appropriate for detecting light dark matter particles in the keV region. We analyze theoretically some key issues involved  in such a detection and  perform calculations for the expected rates employing reasonable theoretical models.
\end{abstract}
\maketitle
\section{Introduction}

The combined earlier results MAXIMA-1 \cite{MAXIMA1},\cite{MAXIMA2},\cite{MAXIMA3}, BOOMERANG \cite{BOOMERANG1},\cite{BOOMERANG2}
DASI \cite{DASI02} and COBE/DMR Cosmic Microwave Background (CMB)
observations \cite{COBE}, \cite{SPERGEL}  imply that the Universe is flat
\cite{flat01}
and that most of the matter in
the universe is Dark \cite{SPERGEL}. These results have been confirmed and improved
by the recent WMAP  \cite{WMAP06} and  Planck data \cite{PlanckCP13}. Combining these data one finds:
$$\Omega_b=0.0456 \pm 0.0015, \quad \Omega _{\mbox{{\tiny CDM}}}=0.228 \pm 0.013 , \quad \Omega_{\Lambda}= 0.726 \pm 0.015~$$
Since, on the other hand, any  non exotic component cannot exceed $40\%$ of the above $ \Omega _{\mbox{{\tiny CDM}}}$
~\cite {Benne}, exotic (non baryonic) matter is required. \\
 On the smaller scales there exists firm indirect evidence from the
observed rotational curves, see e.g. the review \cite{UK01}, for a halo of dark matter
in galaxies and dwarf galaxies. 

Anyway in spite of the above indirect evidence for the existence of dark matter at all scales, it is essential to directly
detect such matter in order to 
unravel the nature of its constituents. 

It clear that the direct detection of dark matter depends on the nature of the dark matter constituents and their interactions.

These, called  WIMP's (Weekly interacting particles), are  expected to have a velocity distribution with an average velocity, close to the rotational velocity $\upsilon_0=220$ km/s of the sun around the galaxy, i.e.  they are completely non relativistic. In fact a Maxwell-Boltzmann distribution with a maximum cut off of about 2.84$\upsilon_0$  leads to a maximum energy transfer close to the average WIMP kinetic  energy  $\prec T\succ\approx 0.4\times 10^{-6}m c^2$. Thus for GeV WIMPS this average is in the KeV regime, not high enough to excite the nucleus, but sufficient to measure the nuclear recoil energy. For light dark matter particles in the MeV region, which we will also  call WIMPs,  the average energy that can be transferred is  in the eV region.

  In the present work we will focus on light WIMPs with a mass less 10 times  the electron mass. So they can be detected  by measuring the  electron recoil, following the WIMP-electron interaction in some targets that  posses weakly bound electrons . Much lighter WIMPs can only be detected by special materials involving very weakly bound electrons, like superconductors by measuring the total deposited energy.

 The event rate for such a process can
be computed from the following ingredients~\cite{LS96}: i) The elementary WIMP-electron cross section. ii)  The WIMP  density in our vicinity obtained from the rotation curves. Due to the assumed smallness of the WIMP mass,  this is expected to be about  six orders of magnitude lager than that involved in the usual WIMPs considered in nuclear recoils. iii) The WIMP velocity distribution.
In the present work we will consider a Maxwell-Boltzmann (MB) distribution in the galactic frame, with the WIMP velocity appropriately  transformed  in  the local frame.

In all  recoil experiments, like the nuclear measurements first proposed more than 30 years ago \cite{GOODWIT}, in order to overcome the formidable  background problems  one can exploit the modulation effect,  a periodic signal due to the motion of the earth around   the sun. Unfortunately this effect, also proposed a long time ago~\cite{Druck} and subsequently studied by many authors~\cite{%
PSS88,GS93,RBERNABEI95,LS96,ABRIOLA98,HASENBALG98,JDV03,GREEN04,SFG06,FKLW11}, 
in the case of nuclear recoils. 

In spite of these problems many experimental undertook the task of detecting nuclear recoils in WIMP-nucleus scattering, see e.g. \cite{XMASS09,CDMSII09,EDELWEISS11,KIMS12,SIMPLE12,PICASSO12,DAMAEPJ13,
CRESST,XENON10017,LUX14}. 
 None has been detected but very stringent limits on the nucleon cross section have been set which can be found in a recent review\cite{KSTH18}. Furtherore projected sensitivities of Dark Matter direct detection experiments to effective WIMP-nucleus couplings have also appeared\cite{LUXZEP}.

The above results combined with theoretical motivations stimulated interest in lower mass WIMPs, see e.g. the recent work \cite{EMV12}. In fact the first direct detection limits on sub-GeV dark matter from XENON10 have recently been obtained \cite{EMMPV12}.   It is, however, clear that  Light WIMPs are quite different in  energy,  mass. 
One, thus,  needs suitable detectors, which maybe completely different from  current WIMP detectors employed for heavy WIMP searches. It is encouraging that light WIMPs in the keV  region can be detected employing Superfluiid Helium  \cite{SchZur16}.

For WIMPs in the mass range of the electron mass, since the available energy is in the eV region, the detection of    electron recoils is possible only for electrons with very low binding energies. Furthermore the detector should be able to measure recoil energy  in few eV region.

Regarding the elementary WIMP-electron cross section we will consider two models:\\
i) Scalar WIMPs, which are viable cold dark matter candidates. Their mass, as far as we know, has not been constrained by any experiment. This scalar WIMP couples with ordinary Higgs with a quartic coupling, which has been inferred  by the LHC experiments. Thus the WIMP interacts  with electrons via Higgs exchange with an amplitude proportional to the electron mass $ m_e$. \\ 
ii)For comparison we will  consider a model with a fermion WIMP interacting via  a Z-exchange with the electron, with a coupling determined phenomenologically. This model, due to the axial coupling, leads to a spin interaction of the electron 

In the present paper  we will address the 
 implications of light scalar WIMPs on the expected event rates scattered off electrons.  
The scalar WIMPs have the characteristic feature that the elementary cross section in their scattering off ordinary quarks or electrons  is increasing as the WIMPs get lighter, which leads to an interesting experimental feature, provided, of course, that the low energy electrons can be detected. For comparison we will also  consider light Fermion WIMPs interacting with the electrons via Z-exchange.

The paper is organized as follows: In section  \ref{sec:particlmodel} we discuss the particle model employed. In section \ref{sec:freeelectrons} we study the detection of essentially free electrons in special low temperature detectors, e.g. superconducting materials, which act as caloremeters. We eill exploit the enhancent of the obtained rates due to the scalar nature of the WIMPs. In section \ref{sec:boundelectrons} we discuss the effect  of the electron binding on the expected rates  in the case of  experiments measuring electron recoils\footnote{We will not concern ourselves here with    two-dimensional targets like those considered recently, see e.g. \cite{HKLTZ17},\cite{DEMSY17}. Such detectors will be considered separately elsewhere \cite{KopVer19}.} in the case of WIMPs with a mass a bit higher than that of the electron. In section \ref{sec:atomicexcitations} we discuss the possibility  of detecting light WIMPs via atomic excitations. This can occur via the spin induced atomic transitions with excitation energy much smaller than the electron binding energy.
\section{The particle model.}
\label{sec:particlmodel}
We will consider two such models:
\subsection{Scalar WIMPs interacting with the Higgs particle in a quartic coupling.}
Scalar WIMP's can occur in particle models. Examples are i) In Kaluza-Klein theories for models involving    universal extra dimensions (for applications to direct dark matter detection  see, e.g.,~\cite{OikVerMou}). In such models  the scalar WIMPs are characterized by ordinary couplings, but they are expected to be quite massive. ii) extremely light  particles ~\cite{Fayet03}, which are not relevant to the ongoing WIMP searches ii) Scalar WIMPs such  as those  considered previously in various extensions of the standard  model~\cite{Ma06}, which  can be quite  light and long lived protected by a discrete symmetry.

Here we will consider as  WIMP  a scalar  particle $\chi$  interacting with another scalar $\phi$, e.g. the Higgs scalar, via a quartic coupling \cite{ZeeScal85,ZeeScal01,BentoRos01,BentoBero00}, and more recently  \cite{Cheung:2014pea}.  
The interest in such a WIMP has recently been revived due to a new scenario of dark matter production in bounce cosmology~\cite{Li:2014era, Cheung:2014nxi} in which the authors point out the possibility of using dark matter as a probe of a  big bounce at the early stage of cosmic evolution. 
A model independent study of dark matter production in the
contraction and expansion phases of the Big Bounce reveals a new venue for achieving the observed relic abundance in which dark matter was produced completely out of chemical equilibrium\cite{Cheung:2014pea} . 
In this way, this alternative route of dark matter production in bounce cosmology can be used to test the bounce cosmos hypothesis \cite{Cheung:2014pea}.
  
\begin{figure}[!ht]
\begin{center}
\subfloat[]
{
\includegraphics[width=0.8\textwidth]{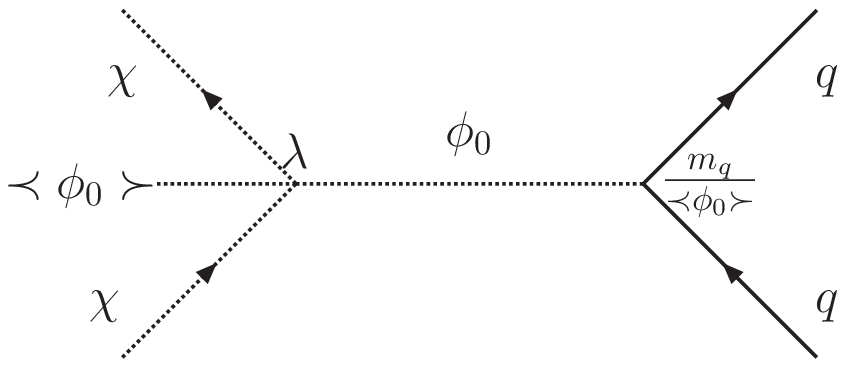}
}\\
\subfloat[]
{
\includegraphics[width=0.8\textwidth]{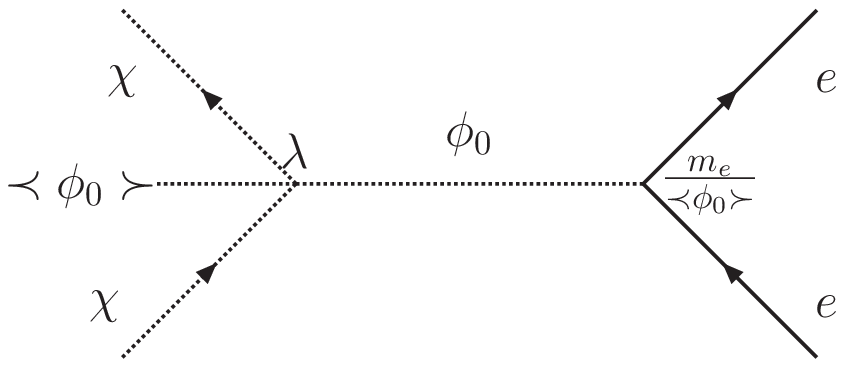}
}
\caption{(a) The quark - scalar WIMP  scattering mediated by a scalar particle. Note that the amplitude is independent of the vacuum expectation value $\prec\phi_0\succ=v$ of the scalar. (b) The corresponding diagram for electron scalar-WIMP
scattering.}
 \label{fig:xxphiphiqe}
 \end{center}
  \end{figure}

In fact the quartic coupling
 \beq
 \phi+\phi \rightarrow \chi+\chi
 \eeq 
involving the scalar WIMP $\chi$ and the Higgs  scalar $\phi=h$ discovered at LHC,  leads to the Feynman diagram shown in Fig. \ref{fig:xxphiphiqe}. 
 In the case of   the  proton the cross section has previously been discussed \cite{Cheung:2014pea}. In the case of the electron the elementary   cross section is
\beq
\sigma= \lambda^2 \frac{1}{(2 m_{\chi})^2} \frac{m_e^2}{ m^4_H}\frac{1}{2 \pi} 2 \mu^2_r=\sigma_{0H}\frac{1}{(1+x)^2}.
\eeq
or
\beq 
\sigma_{0H}=\frac{1}{4 \pi}\lambda^2 \frac{m_e^2}{m_H^4}=8.4\times 10^{-45}\mbox{ cm}^2= 8.4\times 10^{-9}\mbox{pb}
\eeq
 In deriving this scale we have assumed that the quantity $\lambda$ is the same with the quartic coupling appearing in the Higgs potential. This is determined by the LHC data, $\lambda=1/2$.  In the context of dark matter interactions this is  a rather large cross section. It is the result of the fact that, in the small Yukawa coupling $f=\frac{m_e}{v}$, the vacuum expectation value $v$ is canceled by that appearing n the quartic coupling. We thus emphasize that  the cross section  does not suffer from the suppression expected in the decay $h\rightarrow e^{-}e^{+}$ in which $f$ appears and, thus, it cannot be constrained by the LHC data. To the best of our knowledge it is not constrained by any other data.
\subsection{Fermion WIMPs interacting via Z-exchange.}	
Such a mechanism has been considered in the case of the lightest supersymmetric particle (LSP) for the spin induced  hadron cross section and more recently in the WIMP electron scattering \cite{VMCEKL18}. The resulting cross section depends on the coupling of the dark neutral fermions to the Z-boson, i.e. it depends on the nature of the standard model (SM) fermion and the nature of the dark matter:
\beq
{\cal L}= \frac{1}{2 \sqrt{2}}G_F J_{\mu}^Z(\chi)J^{z\mu}(f)= \frac{1}{2 \sqrt{2}}G_FK_{\mu}(\bar{f}\gamma^{\mu}(g_V-g_A\gamma_5)f
\eeq
We are interested in the axial current component, since the Fermi-like coupling of the electron vanishes. We will assume that axial current coupling of the WIMP is also unity. $K=g_{chi}=1$. Then the invariant amplitude squared takes the form:
$${\cal M}^2=\frac{1}{8}G^2_Fg_A^2$$
Proceeding as in the previous subsection we find
\beq
d \sigma= \frac{1}{\upsilon} \frac{1}{8}G^2_F  q^2 dq d \xi \delta(q\upsilon \xi-\frac{q^2}{2\mu_r}),\,\mu_r=\mbox{reduced mass of the WIMP electron system}, 
\eeq
which leads to the total cross section:
\beq
\sigma_e= \frac{1}{8}G^2_F \frac{1}{\pi}\mu_r^2= \frac{1}{8}G^2_F \frac{1}{\pi}m_e^2\frac{x^2}{(1+x)^2}=\sigma_{0Z}\frac{x^2}{(1+x)^2}
\eeq
with
\beq
\sigma_{0Z}=\frac{1}{8}G^2_F \frac{1}{\pi}m_e^2\approx1.0 \times 10^{-9}
\eeq

One may try to infer the electron  cross section from information on the the corresponding  the nucleon cross section.  In fact 
this cross section has been constrained by the  WIMP-nucleus scattering for a WIMP mass, e.g.  of 2 GeV, i.e. $\mu_r=\frac{2}{3}m_p$, by the CRESST-TUM40 experiment \cite{CRESSTTUM40} to be $5\times 10^{-3}$ pb. 

Using the above constrain we obtain:
\beq
\frac{\sigma_e}{\sigma_N}=\frac{9}{4}\frac{m_e^2 m_{\chi}^2}{m_p^2(m_e+m_{\chi})^2}\Rightarrow 
\sigma_e=\sigma_{0Z}\frac{x^2}{(1+x)^2},\,x=\frac{m_e}{m_{\chi}}
\eeq
with
$$\sigma_{0Z}=\frac{9}{4}\frac{m_e^2}{m_p^2}\sigma_N=3.2\times 10^{-45}\mbox{ cm}^2\approx3.2\times 10^{-9}\mbox{ pb} $$
 This value is a factor of 3 larger than the elementary cross section obtained above. Both of them are a bit smaller compared to  the   value $\sigma_{0}=5.0\times 10^{-9}$pb determined phenomenologically \cite{VMCEKL18}. All of them are smaller than that associated with the scalar WIMP obtained above.
  
In this work we will assume for  simplicity common elementary cross section $\sigma_{0}$,
\beq
\sigma_{0H}\approx\sigma_{0Z}=\sigma_{0}=4.0\times 10^{-45}\mbox{cm}^2=4.0\times 10^{-9}\mbox{pb}
\eeq

	\section{The  WIMP-electron rate for free electrons}
	\label{sec:freeelectrons}
	The evaluation of the rate proceeds as in the case of the standard WIMP-nucleon scattering, but we will give the essential ingredients here to establish notation. We will begin by examining the case of a free electron.
 i) The case of the scalar WIMPs (SW):\\
The differential cross when all particles involved are non relativistic and the initial electron is at rest can be cast in the form:
	
	\beq
d \sigma=\frac{1}{\upsilon} \lambda ^2 \frac{1}{(2 m_{\chi})}^2 \frac{m_e^2}{ m_H^4}\frac{1}{(2 \pi)^2} d^3{\bf p}'_{\chi}d^3{\bf q} \delta({\bf p}_{\chi}-{\bf p}'_{\chi}-{\bf q})\delta\left (\frac{{\bf p}^2_{\chi}}{2 m_{\chi}}-\frac{{\bf p}'^2_{\chi}}{2 m_{\chi}}-\frac{{\bf q}^2}{2m}\right )
\eeq
where the factor $1/(2 m_{\chi})^2$ the usual normalization for the scalar particles and $m_H\approx 126$ GeV the mass of the exchanged Higgs particle. Integrating over the momenta we find:
\beq
d \sigma=\frac{1}{2}\sigma_{0H}\frac{1}{ m_{\chi}^2} \frac{1}{\upsilon} q^2 dq d \xi \delta(q\upsilon \xi-\frac{q^2}{2\mu_r}),\,\mu_r=\frac{m_e m_{\chi}}{m_e+m_{\chi}}=\mbox{reduced mass}, 
\eeq
 
	From the energy conserving $\delta$ function one finds  tat the momentum ${\bf q}$ transferred to the electron  is given by $$q=2 m_r \upsilon,\,\upsilon=\mbox{ WIMP velocity},\, \xi=\hat{\upsilon}.\hat{q}\geq 0$$
	Integrating over $\xi$ with the use of the delta function one finds :
	\beq
	d\sigma=\sigma_0\frac{1}{\upsilon^2}\frac{1}{2 m_{\chi}^2 } m_e dT=\sigma_0\frac{1}{2\upsilon^2}\frac{1}{x^2}\frac{dT}{m_e},\,x=\frac{m_{\chi}}{m_e},\,\sigma_0=\sigma_{0H}
	\label{Eq.difsigma}
	\eeq
	Where $T$ is the kinetic energy of the outgoing electron given by:
	\beq
	T=\frac{q^2}{2m_e}=2 \frac{1}{m_e}\mu^2_r \upsilon^2 \xi^2=2 m_e\frac{m_{\chi}^2 }{m_e^2+m_{\chi}^2 }\upsilon^2 \xi^2=2 m_e\upsilon^2 \xi^2\frac{x^2}{(1+x)^2}
	\label{Eq:Ttransf}
	\eeq
	
ii) The case of the fermion WIMP (FW).\\
Proceeding as above we find
	\beq
	d\sigma=\sigma_{0Z}\frac{1}{2\upsilon^2} \frac{dT}{m_e}
	\label{Eq.difsigma1}
	\eeq

	From Eq. (\ref{Eq:Ttransf}) , after integrating over the angles, we find  that the fraction of the energy  of the WIMP  transferred to the electron  is
	\beq
	\frac{T}{T_{\chi}}=4 \frac{x}{(1+x)^2},\,x=\frac{m_{\chi}}{m_e}
	\label{Eq:Ttransfratio}
	\eeq
	We thus see that this ratio becomes unity, i.e. maximum, when $x=1$. 

\begin{figure}[!ht]
  \begin{center}
\includegraphics[width=0.7\textwidth,height=0.4\textwidth]{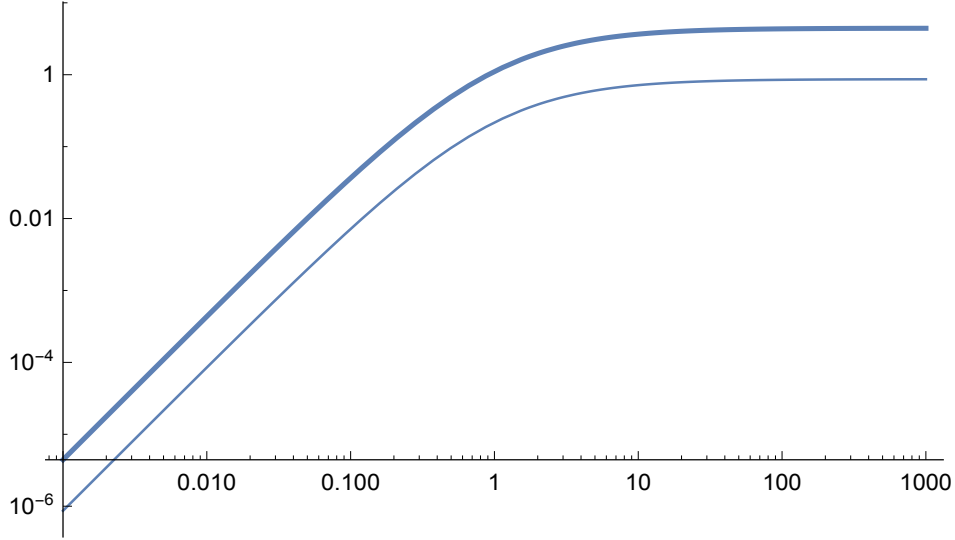}\\
 \caption{The  maximum (thick solid line) and the average (fine solid line) energy transfer in eV  as a function of $x=\frac{m_{\chi}}{m_e}$ in the case of a free electron.}
 \label{fig:TmaxTav}
 \end{center}
  \end{figure}

 The maximum energy transfer depends on the escape velocity, which is assumed to be  $\upsilon_{esc}\approx 2.84 \upsilon_0$ with  $\upsilon_0=0.7 10^{-3}c$ the sun's velocity round the center of the galaxy. Integrating  the energy transfer  over the velocity distribution we obtain the average energy transfer. The maximum and the average energy transfer are exhibited in fig. \ref{fig:TmaxTav}.
 Thus for MeV WIMP the average energy transfer is in the eV region, which is reminiscent of the standard WIMPs where GeV mass leads to an energy transfer in the keV region. The same  the average energy is obtained by the convolution the  energy transfer with the differential rate, which will be given below (for more details see \cite{VMCEKL18}).

	 Furthermore for a given energy transfer $T$ we find:
	\beq
	\upsilon=\sqrt{\frac {m_eT}{2 \mu^2_r\xi^2}}=\left(1+\frac{1}{x}\right)\sqrt{\frac{1}{2} \frac{T}{m_e}}\rightarrow \upsilon\ge\left(1+\frac{1}{x}\right)\sqrt{\frac{1}{2} \frac{T}{m_e}}\rightarrow \upsilon_{min}=\left(1+\frac{1}{x}\right )\sqrt{\frac{1}{2} \frac{T}{m_e}}.
	\eeq
In other words the minimum velocity consistent with the energy transfer $T$ and the WIMP mass is constrained as above. The maximum velocity allowed is determined by the velocity distribution and it will be indicated by $\upsilon_{esc}$.
From this we can obtain the differential rate per electron in a given velocity volume $\upsilon^2 d \upsilon d \Omega$ as follows:
	\beq
	dR=\sigma_0\frac{\rho_{\chi}}{m_{\chi}}\frac{1}{2}\upsilon \nu(x)\frac{dT}{m_e} f({\vbf}) d\upsilon d\Omega,\,\nu(x)=\left \{ \begin{array}{cc}\frac{1}{x^2}, &\mbox{{\tiny SW}}\\ 1,&\mbox{{\tiny FW}}\\ \end{array} \right .
	\eeq
	where $ f({\vbf})$ is the velocity distribution of WIMPs in the laboratory frame. Integrating over the allowed velocity distributions we obtain:
	\beq
	dR=\frac{\rho_{\chi}}{m_{\chi}}  \sigma_0 \frac{dT}{m_e}\frac{1}{2\upsilon_0} \eta(\upsilon_{\mbox{\tiny{min}}})\times \left \{ \begin{array}{cc}{x^2}, &\mbox{{\tiny SW}}\\ 1,&\mbox{{\tiny FW}}\\ \end{array} \right .,\,\eta(\upsilon_{\mbox{\tiny{min}}})=\int_{\upsilon_{\mbox{\tiny{min}}}}^{\upsilon_{esc}} f({\vbf})\upsilon d\upsilon d\Omega
	\label{Eq:elrate1}
	\eeq
	The parameter $\eta(\upsilon_{\mbox{\tiny{min}}})$ is a crucial parameter.\\
	Before proceeding further we find it convenient to express the velocities in units of the Sun's velocity. We should also take note of the fact the velocity distribution is given with respect to the center of the galaxy. For a M-B distribution this takes the form:
	\beq
	\frac{1}{\pi \sqrt{\pi}}e^{- y^{'2}},\,y^{'}=\frac{\upsilon^{'}}{\upsilon_0},\,\upsilon_0=220 \mbox{ km/s}
	\eeq
	We must transform it to the local coordinate system :
	\beq
	{\bf y}^{'}\rightarrow {\bf y}+{\hat\upsilon}_s+ \delta \left
(\sin{\alpha}{\hat x}-\cos{\alpha}\cos{\gamma}{\hat
y}+\cos{\alpha}\sin{\gamma} {\hat \upsilon}_s\right ) ,\,\delta=\frac{\upsilon_E}{\upsilon_0}
 \label{Eq:vlocal} \eeq 
with
$\gamma\approx \pi/6$, $ {\hat \upsilon}_s$ a unit vector in the
Sun's direction of motion, $\hat{x}$  a unit vector radially out
of the galaxy in our position and  $\hat{y}={\hat
\upsilon}_s\times \hat{x}$. The last term,  in parenthesis, in
 Eq. (\ref{Eq:vlocal}) corresponds to the motion of the Earth
around the Sun with $\upsilon_E\approx 28$ km/s being  the modulus of the
Earth's velocity around the Sun and $\alpha$ the phase of the Earth ($\alpha=0$ around June 3nd). The above formula assumes that the
motion  of both the Sun around the Galaxy and of the Earth around
the Sun are uniformly circular. Since $\delta$ is small we can expand the distribution in powers of $\delta$ keeping terms  up to linear in $\delta$.
\beq
			dR=\left (\frac{\rho_{\chi}}{m_{\chi}} {\upsilon_0}\right ) N_e \frac{1}{2 \upsilon_0^2}\frac{dT}{m_e}\left ( \Psi_0(y_{min})+
	 \Psi_1(y_{min}) \cos{\alpha} \right )\times \left \{ \begin{array}{cc}{\frac{1}{x^2}}, &\mbox{{\tiny SW}}\\ 1,&\mbox{{\tiny FW}}\\ \end{array} \right .,\,x=\frac{m_{\chi}}{m_e},
	\label{Eq:elrate2}
	\eeq
where in the above equation the first term in parenthesis represents the average  flux of WIMPs, the second   term gives the number $N_e$ of electrons available for the scattering \footnote{In standard targets $N_e=\frac{m_t Z_{eff}}{A m_p}$, in a target of mass $m_t$ containing atoms with mass number $A$, $Z_{eff}$ represents the number of available electrons. The meaning of $Z_{eff} $ becomes clear if one takes into account that the electrons are not free but bound in the atom 
see section \ref{sec:boundelectrons}. Thus they are not all available for scattering, i.e.  $Z_{eff}<<Z $.}:
. 
Furthermore  for a M-B distribution one finds \cite{VMCEKL18}:
\beq
\Psi_0(x)=\frac{1}{2}H\left (y_{esc}-x \right )
  \left [\mbox{erf}(1-x)+\mbox{erf}(x+1)+\mbox{erfc}(1-y_{\mbox{\tiny{esc}}})+\mbox{erfc}(y_{\mbox{\tiny{esc}}}+1)-2 \right ],\,x=y_{\mbox{\tiny min}}
  \label{Eq:Psi0MB}
\eeq
and
\barr
\Psi_1(x)&=&\frac{1}{2} H\left (y_{esc}-x \right )\delta 
   \left[\frac{ -\mbox{erf}(1-x)-\mbox{erf}(x+1)-\mbox{erfc}(1-y_{\mbox{\tiny{esc}}})-
   \mbox{erfc}(y_{\mbox{\tiny{esc}}}+1)}{2} \right . \nonumber\\
  && \left . +\frac{ e^{-(x-1)^2}}{\sqrt{\pi }}
   +\frac{
   e^{-(x+1)^2}}{\sqrt{\pi }}-\frac{ e^{-(y_{\mbox{\tiny{esc}}}-1)^2}}{\sqrt{\pi
   }}-\frac{ e^{-(y_{\mbox{\tiny{esc}}}+1)^2}}{\sqrt{\pi }}+1\right],\,x=y_{\mbox{\tiny min}}
   \label{Eq:Psi1MB}
\earr
with
$$
y_{min}=\frac{\upsilon_{min}}{\upsilon_0}=\frac{1}{\upsilon_0}\left (1+\frac{1}{x}\right)\sqrt{\frac{1}{2} \frac{T}{m_e}},\,y_{esc}=\frac{\upsilon_{esc}}{\upsilon_0}
$$
In the above expression  the Heaviside function $H$ guarantees that the required kinematical condition is satisfied. 
After this we are going to proceed in evaluating the expected spectrum of the recoiling electrons.

The expression given by Eq. (\ref{Eq:elrate2} ) can be cast in the form:
\beq
\frac{dR}{d(T/1\mbox{eV})}=\rho\Lambda \left (\Sigma_0\left (\frac{m_{\chi}}{m_e},\frac{T}{\left(1\mbox{eV}\right )}\right )+\Sigma_1\left (\frac{m_{\chi}}{m_e},\frac{T}{\left (1\mbox{eV}\right )}\right ) \cos{\alpha} \right ),\rho=\frac{1 \mbox{eV}}{2 m_e\upsilon_0^2}\approx2
\label{Eq:ediffRate}
\eeq
where 
\beq
\Sigma_i(x,s)=\frac{1}{x}\Psi_i\left (1.23\left(1+\frac{1}{x}\right )\sqrt{\rho s}\right )\times \left \{ \begin{array}{cc}{\frac{1}{x^2}}, &\mbox{{\tiny SW}}\\1,& \mbox{{\tiny FW}}\\\ \end{array} \right . ,\,i=0,1,\,s=\frac{T}{1\mbox{eV}}
\label{Eq:Sigma}
\eeq
and
\beq
	\Lambda=\frac{\rho_{\chi}}{m_e} \sigma_0 {\upsilon_0} N_e 
	\label{Eq:elrate3}
	\eeq
	Where $N_e$ the number of electrons in the target.
	
\begin{figure}[!ht]
  \begin{center}
	\subfloat[]
	{
\includegraphics[width=0.7\textwidth,height=0.3\textwidth]{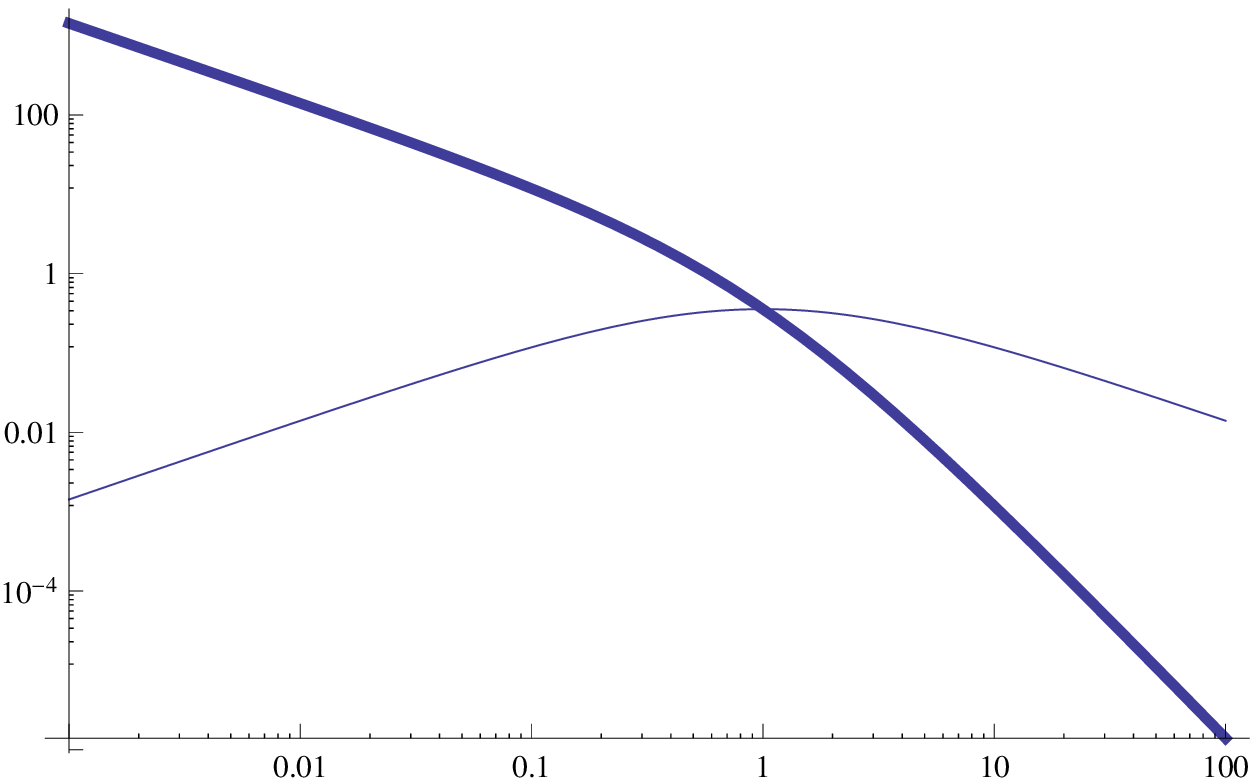}
}\\
	\subfloat[]
	{
\includegraphics[width=0.7\textwidth,height=0.3\textwidth]{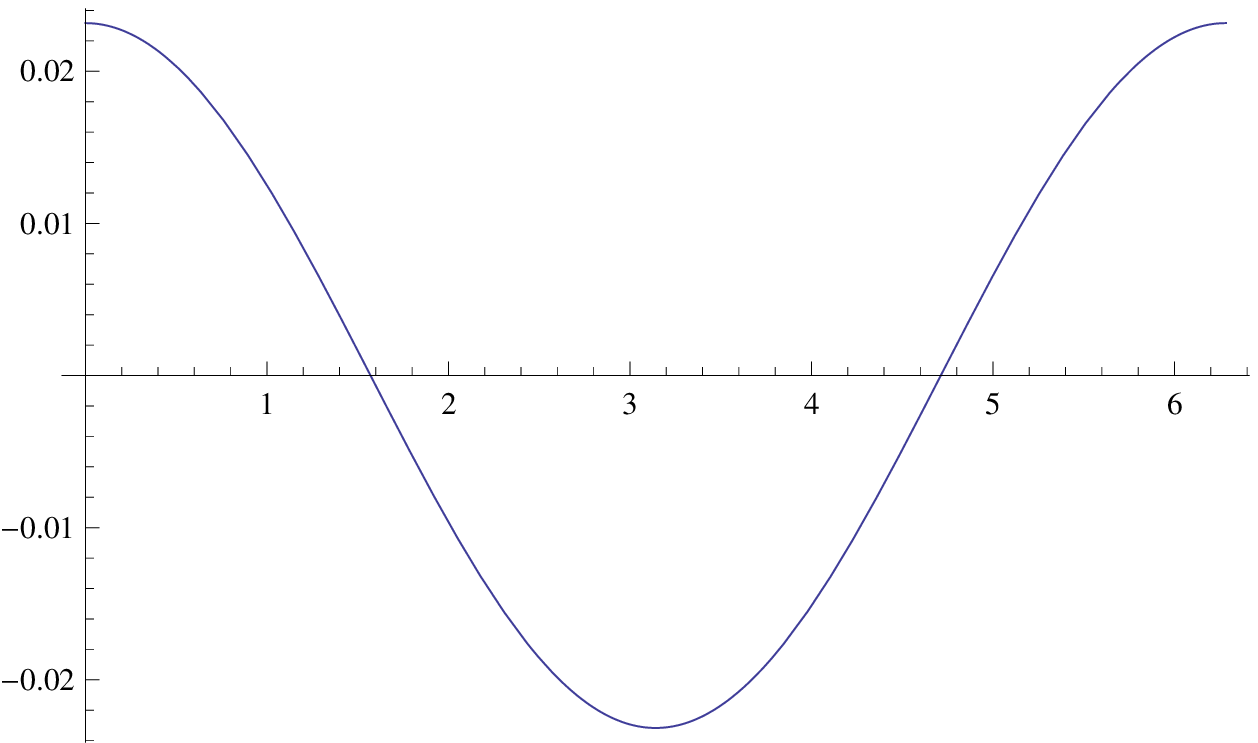}
}
 \caption{(a) The total  time averaged event rate $R_0$,  in units of $\Lambda$ , as a function of  $x=\frac{m_{\chi}}{m_e}$. The thick   and the fine solid lines correspond to a scalar and Fermion WIMP respectively.(b)  The ratio of the time dependent to the time average rate,$\frac{R_1}{R_0}\cos{\alpha}$, as a function of the phase of the Earth $\alpha$ ($\alpha=0$ around June 3nd).}
 \label{fig:totrate}
 \end{center}
  \end{figure}

	The total  event  rates are  given by:
	\beq
	R_i=\Lambda \rho \int_0^{s_{\mbox{\tiny{max}}}}\Sigma_i(x,s),\,s_{\mbox{\tiny{max}}}=\frac{T_{\mbox{\tiny{max}}}}{\mbox{1 eV}}, \, \frac{1}{2 \upsilon_0^2}\approx 10^{6} .
	\eeq
	The time average rate $R_0$  is exhibited in Fig. \ref{fig:totrate}a.\\ For the time dependence we prefer to present:
	\beq
	R_r=\frac{R_1}{R_0} \cos{\alpha},\, \alpha= \mbox{the phase of the Earth},
	\eeq
	Where $R_r$ is essentially independent of $x$ and is exhibited in Fig. \ref{fig:totrate}b.

	It is thus obvious for light WIMPs it is necessary to consider special materials in which the electrons are loosely bound, like electron pairs in a superconductor, provided, of course, that the number of these electrons is not very small.\\ We will now estimate the rate for free electrons, i.e.  estimate $\Lambda$ considering the following input:.


	\begin{itemize}
	\item the elementary cross section $\sigma_{0}=4\times 10^{-9}pb=4 \times 10^{-45}\mbox{cm}^2$ both for the Z and Higgs exchage.
	\item The particle density of WIMPs in our vicinity:
	$$n =0.3 \times10^3 \mbox{(MeV /cm}^3\mbox{)/0.511MeV}\approx 600\mbox{cm}^{-3}$$
	(we use the electron mass in this estimate, since  the correct mass dependence has been included through the extra factor of $x$ in   Eq. (\ref{Eq:Sigma})). This value leads to a flux:
	$$\Phi_0=n\times 220 \mbox{ km/s}=1.3 \times 10^{10}\mbox{cm}^{-2}\mbox{s}^{-1}=4.2 \times 10^{17}\mbox{cm}^{-2}\mbox{y}^{-1}$$
	\item The number of electrons  in the target,  estimated to be
	$$N_e=10^{24}$$
	\end{itemize}
	We thus using Eq. (\ref{Eq:elrate3}) we obtain
	$$ \Lambda\approx 1.7 \times 10^{-3}\mbox{y}^{-1}$$ 
	 From Fig. \ref{fig:totrate}a we find:\\ 
	\begin{itemize}
	\item $x=1\Rightarrow$\\
	 $$R=0.36 \times \Lambda=6. 0 \times 10^{-4}\mbox{y}^{-1}$$ 
	both for Fermion and scalar WIMPs. Maximum for Fermion WIMPs
	\item For scalar WIMPs \\
	$$x= 10^{-2}\Rightarrow R=1.2 \times 10^2 \times \Lambda=0.2\mbox{y}^{-1}$$
	$$x= 10^{-3}\Rightarrow R=1.2 \times 10^3 \times \Lambda=2\mbox{y}^{-1}$$
	\end{itemize}
	
	We should mention, however, that the WIMP detection in calorimetric experiments is still difficult, since, in spite of the large rate in the case of scalar WIMPs, the  total amount energy deposited in the detector for such a light WIMP is very small. 
	
	Anyway it is  encouraging that it seems possible,  as it has  recently been suggested \cite{HPZ15},  to detect even very light WIMPS, much lighter than the electron, utilizing Fermi-degenerate materials like superconductors at low temperatures. In this case the energy required is essentially the gap energy of about $1.5 kT_c$ which is in the meV region, i.e the electrons are essentially free. These authors claim that in spite of the small energy   in the range of few meV deposited to the system, the  detection of very light WIMPs becomes feasible.

	\section{The  WIMP-electron rate for bound electrons}
	\label{sec:boundelectrons}
	In the presence of bound electrons the WIMP mass must be around the mass of the electron, $x=\frac{m_{\chi}}{m_e}\ge 1$. In this case it is advantageous to consider  the $Z$-exchange. Thus the
	 differential cross section for bound electrons \footnote{ Since as we have seen in section \ref{sec:particlmodel}
	 $$\sigma_e=\sigma_{0Z} \frac{x^2}{(1+x)^2}$$
	 \beq
	 \frac{\sigma_{0Z}}{m_e^2}\leftrightarrow\frac{\sigma_{0Z}}{m_e^2}\left(1+\frac{1}{x}\right )^2\leftrightarrow\frac{\sigma_{e}}{\mu^2_r}
	 \eeq
	 The latter form are preferred of the WIMP-electron cross section is determined phenomenologically
	 }
	  takes the form:
	 \beq
	 d \sigma=\frac{\pi}{m_e^2}\frac{1}{\upsilon}\sigma_{0Z}\left |{\cal M}({\bf q})\right |^2\frac{d^3{\bf q}}{(2\pi)^3}\frac{d^3 {\bf p}'_{\chi}}{(2\pi)^3}\frac{d^3 {\bf p}_A}{(2\pi)^3} (2\pi)^3\delta \left ({\bf p}_{\chi}-{\bf p}'_{\chi}-{\bf q} -{\bf p}_A\right ) (2 \pi)\delta \left ( \frac{{\bf p}^2_{\chi}}{2 m_{\chi}}- \frac{({\bf p}')^2_{\chi}}{2 m_{\chi}}-\frac{{\bf }q^2}{2 m_e}\right )
	 \eeq
	 where 
 ${\bf p}_{\chi}$ and ${\bf p}'_{\chi}$ are  the momenta of the oncoming and outgoing WIMPs with mass $ m_{\chi}$ and $\upsilon$ is the velocity of the oncoming WIMP. Further more	 $${\cal M}({\bf q})=\int d{\bf r}e^{i {\bf q}.{\bf r}}\psi_{n_r,\ell,m}$$
	 with $\psi_{n_r,\ell,m}$ the bound electron wave function coordinate  space. ${\cal M}({\bf q})$ essentially represents  
	 the overlap between the  electron bound wave function and the plane wave of the outgoing electron with momentum ${\bf q}$. It can be written as  $(2 \pi)^{3/2}\Phi_{n_r,\ell,m}(\bf{a,q})$,
    with $\Phi_{n_r,\ell,m}(\bf{a,q})$ the bound electron wave function in momentum space. For $\ell=0$ (s-states), which are of interest in the present work, they appear in  table \ref{tab:belwf}.\\ Note that the energy of the atom is negligible and does not appear in the energy conserving $\delta$ function.
\begin{table}
\begin{center}
\caption{ The $\ell=0$ bound electron  wave functions in momentum space. $a=\frac{\alpha Z}{n_r+\ell+1}\frac{m_e c^2}{\hbar c}$, $\alpha\approx\frac{1}{137}$.
\label{tab:belwf}}
$$
\begin{array}{|c|c|}
\hline
n_r&\Phi_{nr}(α,k)\\
0&\frac{2 \sqrt{2} a^{5/2}}{\pi 
   \left(a^2+k^2\right)^2}\\
1&\frac{4 \sqrt{2} a^{5/2}
   \left(k^2-a^2\right)}{\pi 
   \left(a^2+k^2\right)^3}\\
2&\frac{2 \sqrt{2} a^{5/2} \left(3
   a^5-10 a^3 k^2+3 a
   k^4\right)}{\pi
   \left(a^2+k^2\right)^4}\\
3&\frac{8 \sqrt{2} a^{5/2}
   \left(-a^6+7 a^4 k^2-7 a^2
   k^4+k^6\right)}{\pi 
   \left(a^2+k^2\right)^5}\\
4&\frac{2 \sqrt{2} a^{5/2} \left(5
   a^4-10 a^2 k^2+k^4\right)
   \left(a^4-10 a^2 k^2+5
   k^4\right)}{\pi 
   \left(a^2+k^2\right)^6}\\
5&\frac{4 \sqrt{2} a^{5/2} \left(-3
   a^{10}+55 a^8 k^2-198 a^6
   k^4+198 a^4 k^6-55 a^2 k^8+3
   k^{10}\right)}{\pi 
   \left(a^2+k^2\right)^7}\\
	\hline
	\end{array}
$$
\end{center}
\end{table} 

\begin{figure}[!ht]
  \begin{center}
\rotatebox{90}{\hspace{-0.0cm} $\psi^2_{n_r,\ell}(u(T))\rightarrow$}
\includegraphics[width=0.7\textwidth,height=0.4\textwidth]{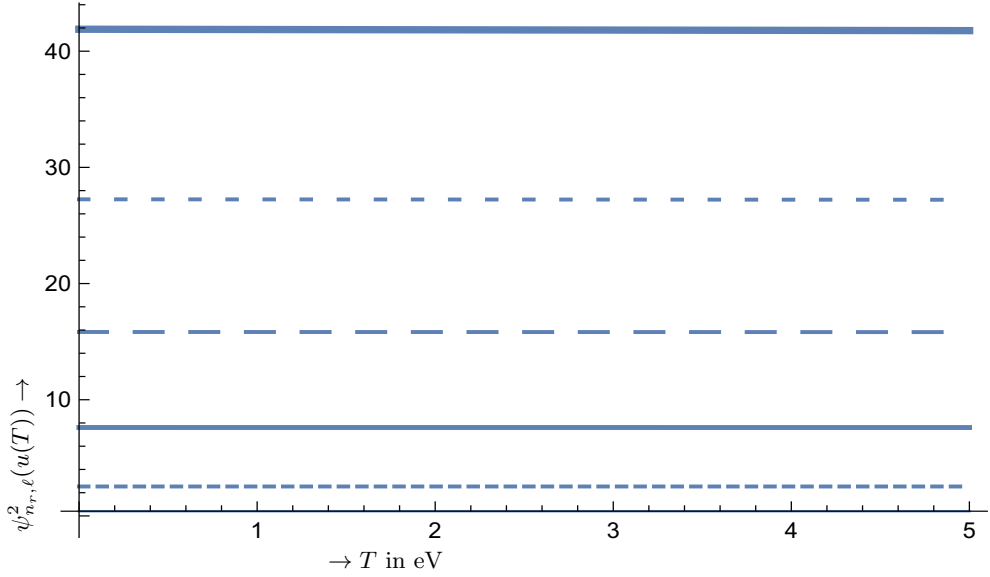}\\
\hspace{-3.0cm}$\rightarrow T$ in eV \\
 \caption{The function $\psi^2_{n_r,\ell}(u(T))$ for $\alpha Z=1/2$ is exhibited as a function of the electron recoil energy $T$ in units of eV. It is shown for $n_r=0,1,2,3,4,5$ increasing upwards (the lowest one is barely visible).}
 \label{fig:phielT}
 \end{center}
  \end{figure}

	 Thus integrating over ${\bf p}_A$ with the help of the momentum conserving $\delta$ function we obtain
	 \beq
	 d \sigma=\frac{\pi}{\upsilon}\frac{\sigma_{0Z}}{m_e^2}\frac{1}{(2 \pi)^2}( \Phi_{n_r,\ell}^2(\bf{a,{\bf q})d^3 {\bf p}'_{\chi}d^3 {\bf q}} \delta \left ( \frac{{\bf p}^2_{\chi}}{2 m_{\chi}}- \frac{({\bf p}')^2_{\chi}}{2 m_{\chi}}-\frac{{\bf }q^2}{2 m_e}\right )
	 \eeq 
	 Then 
	$$\int d^3 {\bf p}'_{\chi}\delta \left ( \frac{{\bf p}^2_{\chi}}{2 m_{\chi}}- \frac{({\bf p}')^2_{\chi}}{2 m_{\chi}}-\frac{{\bf }q^2}{2 m_e}\right )=4 \pi m_{\chi}^2 \upsilon\sqrt{1-\frac{2(b+T}{m_x\upsilon^2}} $$ 
	where $ T$ is the recoiling energy of the electron $T=q^2/(2 m_e)$. Similarly  the integration over ${\bf q}$ for s-wave functions yields $ \Phi^2_{n_r,\ell}( a,\sqrt{2 m_e T}) 4 \pi \sqrt{2 m_e T} m_e dT$.
	Furthermore by writing $\sqrt{2 m_eT}=u a$ we get 
	$$\Phi^2_{n_r,\ell}( a,\sqrt{2 m_e T})=\frac{\psi^2_{n_r,\ell}(u)}{a^3}$$
	  Thus the cross section becomes
	$$
	d \sigma=\frac{4 \pi}{y}\sigma_{0Z} x^2 \psi^2_{n_r,\ell}(u) u \sqrt{y^2-\frac{2(b+T}{x m_e\upsilon_0^2}}\frac{m_e dT}{a^2},
	$$
	where having in mind to eventually use the Maxwell-Boltzmann (M-B) velocity distribution we have expressed the velocities in units of $\upsilon_0=220$km/s. Measuring now the $b$ and $T$ in eV, which is the expected scale we obtain
	\beq
	d \sigma=\frac{4\pi}{y}\sigma_{0Z} x^2\psi^2_{n_r,\ell}(u(T)) 
	\sqrt{y^2-\frac{\rho'(b+T)}{x}}  \left (\frac{n_r+\ell+1}{\alpha Z}\right )^2 \times 2 \times ^{-6}u(T)dT,\, \rho'=3.64
	\label{Eq:dsigmau}
	\eeq
	where
	\beq
	 u(T) = \sqrt{\frac{0.2}{m_e}}\frac{n_r+\ell+1}{\alpha Z}\sqrt{T}\approx 6.3 \times 10^{-4}\frac{n_r+\ell+1}{\alpha Z} \sqrt{T}
	 \eeq
	 The behavior of the function $\psi^2_{n_r,\ell}(u(T))$ for $\alpha Z\approx \frac{1}{2}$  for various values of $n_r$ is exhibited in Fig. \ref{fig:phielT}. One can see that  the higher $n_r$ are favored. For a given $n_r$ it is essentially independent of $T$ for recoiling energies of interest to us. 
Returning now to Eq. (\ref{Eq:dsigmau}) we find some very useful 	limits.\\
i) in folding with the velocity distribution we must integrate between $y_{min}=\sqrt{\frac{2 \rho'(b+T)}{x}}$ and $y_{esc}=2.84$\\
ii) for a given $x$ and $b$ the maximum electron energy is 
$$\frac{T_{max}}{\mbox{1 ev}}=\frac{y^2_{esc}x}{2 \rho' }-\frac{b}{\mbox{1 ev}}=1.1 x-\frac{b}{\mbox{1 ev}}$$
Thus for a value of $x=5$ and a binding energy 2.5 eV the maximum electron energy is expected to be 3 eV.\\ 
iii) For a given binding energy $x$ must be at least $x_{min}=0.90 b$

Folding the cross section with the velocity distribution (see Eq. (\ref{Eq:fyxi}) below) including the extra factor of $y$ coming from the flux we obtain:
\barr
	\langle y \frac{d \sigma}{dT}\rangle&=&4\pi\sigma_{0Z} x^2 \psi^2_{n_r,\ell}\left (\frac{n_r+\ell+1}{\alpha Z}\right )^2 \times 2 \times ^{-6}u(T)dT g(x,T,b),\nonumber\\g(x,T,b)
&=&\frac{2}{\sqrt{\pi}}\int_{y_{min}}^{y{esc}}dy y e^{-(1+y^2)}\sinh{2 y}\sqrt{y^2-\frac{\rho'(b+T)}{x}}
	\earr
The total rate can now be cast in the form
\beq
\frac{dR}{dT}=\Lambda R_{d_0},\,R_{d_0}=4\pi x \psi^2_{n_r,\ell}\left (\frac{n_r+\ell+1}{\alpha Z}\right )^2 \times 2 \times ^{-6}u(T) g(x,T,b)
\label{Eq:dratebound}
\eeq
\beq
R=\Lambda R_0,\,R_0=4\pi x\int_0^{Tmax(x,b)} dT \psi^2_{n_r,\ell}\left (\frac{n_r+\ell+1}{\alpha Z}\right )^2 \times 2 \times ^{-6}u(T) g(x,T,b)
\label{Eq:ratebound}
\eeq
where 
$$ \Lambda= \frac{\rho_{\chi}}{m_e}  \sigma_{0Z} \upsilon _0$$	
with  $\rho_{\chi}$ the WIMP density in our vicinity. Note that that $m_e$ rather $m_{\chi}$ has been employed in determining the number density of WIMPs with a compensating factor $1/x$  already   incorporated into Eq. \ref{Eq:ratebound}.

There exist few atoms which possess s-state electrons with small binding energies. From atomic data tables \cite{Larkins77,Sevier72,PorFreed78} we found and  list those with $b\le 10$ eV in table \ref{tab.atomicb}. There exist of course states with binding energies smaller than those of the s-states, but, as we have mentioned for light WIMPs they are not going to contribute significantly to the total rate.
\begin{table}
\begin{center}
\caption{Listed are the atoms and the indicated binding energy of the corresponding s-electrons. Only electrons with binding energies less than 10 eV are listed.
\label{tab.atomicb}
}  
\begin{tabular}{|c|c|c|c|c|c|c|c|}
\hline
$_{49}$In: 0.1 eV&$_{11}$Na: 0.7 eV&$_{23}$Al: 0.7 eV&$_{50}$Sn: 0.9 eV&$_{31}$Ga: 1.5 eV&$_{12}$Mg: 2.1 eV&$_{65}$Cd: 2.2 eV&
$_{82}$Pb: 3.1 eV\\$_{31}$Ge: 5.0 eV&$_{3}$Li: 5.3 eV&$_{51}$Sb: 6.7 eV&$_{14}$Si: 7.6 eV&$_{83}$Bi: 8.0 eV&$_{33}$As: 8.5 eV&$_{84}$Po: 9.0 eV\\
\hline
\end{tabular}
\end{center}
\end{table}
It thus apperars that i) NaI (b=0.7 eV in Na) as scintillator and ii)  CdTe (b=2.2 eV in Cd), Ge(Li) (b=5 eV in Ge and Li) and Si (b=7.6 eV) can be used as solid state detectors.

\begin{figure}[!ht]
  \begin{center}
\rotatebox{90}{\hspace{-0.0cm} $R\rightarrow $ events/(kg-y)}
\includegraphics[width=0.7\textwidth,height=0.4\textwidth]{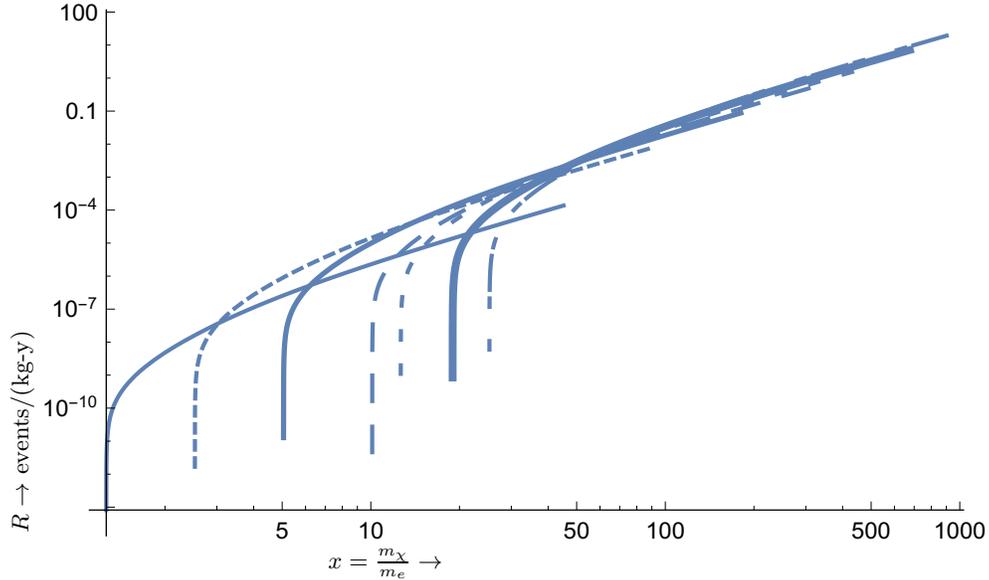}\\
\hspace{-3.0cm}$x=\frac{m_{\chi}}{m_e}\rightarrow $ \\
 \caption{The event rates as a function of $x$ for various electron binding energies, which are  increasing from left to right. Thus the fine solid curve corresponds to $b=1$ $(Z_{eff}=1.0)$, the short dashed curve $b=2$ $(Z_{eff}=1)$, the intermediate thick solid line to $b=3$ $(Z_{eff}=2)$ the long  dashed curve $b=4$ $(Z_{eff}=3)$, the intermediate short dashed curve $b=6$ $(Z_{eff}=4)$, the solid line to  $b=10$ $(Z_{eff}=6)$ and the short-long dashed curve to b=15 $(Z_{eff}=8)$. One can clearly see the threshold values of $x$ for a given binding energy $b$. For illustration purposes the hydrogenic wave function with  $n_r=4,\ell=0$  has been employed. }
 \label{fig:Rate1Zef}
 \end{center}
  \end{figure}
  
  Many of the elements listed in table \ref{tab.atomicb}, involving s-electrons with low binding energies can serve as good targets, provided, of course, that recoiling electrons with energies in the few eV can be detected. Once a special target is selected, one must make an orbit by orbit calculation, based on the data of  table \ref{tab.atomicb}, and sum the cross section over all orbits multiplied with the number of electrons involved.
  
At this point we will make a simple calculation using  $N_e=N_A=10^{25}$, which corresponds to the number of atoms of a Kg of an $A=60$ target and is an order of magnitude larger than that used in the case of free electrons  discussed in the previous section.  We thus obtain the results shown in Fig. \ref{fig:Rate1Zef}  using  $Z_{eff}$ much smaller then $Z$ for a typical atom. In spite of the larger $N_e$,  for low  $x$ the obtained results are  smaller  than those obtained in the previous section. We can trace this suppression to the atomic parameter $a$, which is of the order of $m_e$, much larger than the electron recoiling energies, which, for  $x<10$, tend to be in the few eV region.

   The results, of course, tend to further  increase  approximately linearly with $x$ and eventually, for $x>50$, electron recoils become easily detectable. For such values of $x$, of course, all electrons can participate, i.e. $Z_{eff}=Z$
	\section{Atomic excitations}
	\label{sec:atomicexcitations}
	We have seen that detecting low mass WIMPs by observing  recoiling electrons is pretty hard, since few electrons can be ejected, due to their binding in the atom. This problem does not persist, if the electrons are not ejected, but promoted to a higher level and the de-excitation photons are observed. In this case  an energy difference even much smaller than eV is possible, if the target is placed  in a magnetic field at low temperature.
	
	As a matter of fact the  axial current present in the Z-mediated WIMP-electron interaction through the electron spin  can cause atomic transitions between atomic levels within states, which have the same radial quantum numbers and angular quantum numbers $j_1,m_1$ and $j_2,m_2$. If the atom is placed in a magnetic field the transition matrix element is expressed in terms of the Glebsch-Gordan coefficient and the nine- j symbol:
	\barr
	{\cal M}[(n,\ell,j_1m_1)\rightarrow(n,\ell,j_2m_2)]&=&C_{\ell,j_1,m_1,j_2,m_2},\nonumber\\
	C_{\ell,j_1,m_1,j_2,m_2}&=&\langle j_1\,m_1,1\,m_2-m_1|j_2\,m_2\rangle\sqrt{(2 j_1+1)3}\sqrt{2 \ell+1}\sqrt{6}\left \{ \begin{array}{ccc}\ell&\frac{1}{2}&j_1\\\ell&\frac{1}{2}&j_2 \\0&1&1 \end{array}\right \}
	\earr
	When $j_1=j_2$ the two states are those arising from the splitting of the degeneracy due to the Zeeman effect with an energy difference $\delta E=E_f-E_i= \mbox{ a few}\mu$eV. If $j_1\ne j_2$ the two levels correspond the spin orbit partners with energy differences in the eV region. For the readers convenience these matrix elements are tabulated for some cases of practical interest will be given below.
	
	The differential cross section now takes the form:
	\beq
d \sigma=\frac{1}{\upsilon} \sigma_{0Z}  \frac{\pi}{ m^2_e} \frac{1}{(2 \pi)^2}\left (C_{\ell,j_1,m_1,j_2,m_2}\right )^2 d^3{\bf p}'_{\chi}d^3{\bf q} \delta({\bf p}_{\chi}-{\bf p}'_{\chi}-{\bf q})\delta\left (\frac{{\bf p}^2_{\chi}}{2 m_{\chi}}-\frac{{\bf p}'^2_{\chi}}{2 m_{\chi}}-\delta E\right )
\eeq
where $M$ is the mass of the atom and $q$ the momentum transfer to the atom and $\delta E$ the excitation energy. The recoil energy of the atom is negligible.
 Integrating over the momentum ${\bf q}$ we find:
\beq
d \sigma=\frac{1}{\upsilon} \sigma_{0Z}\frac{\pi}{ m^2_e} \frac{1}{(2 \pi)^2 } \left (C_{\ell,j_1,m_1,j_2,m_2}\right )^2 d^3{\bf p}'_{\chi}\delta\left (\frac{{\bf p}^2_{\chi}}{2 m_{\chi}}-\frac{{\bf p}'^2_{\chi}}{2 m_{\chi}}-\delta E\right ).
\eeq
Performing the remaining integration we get
\beq
d \sigma=\frac{1}{\upsilon} \sigma_{0Z}\frac{\pi}{ m^2_e} \frac{1}{(2 \pi)^2} \left (C_{\ell,j_1,m_1,j_2,m_2}\right )^2 4 \pi \left .\frac{{{\bf p}'^2_{\chi}}}{| p'_{\chi}/m_{\chi}}\right |_{p'_{\chi}=\sqrt{p^2_{\chi}-2 m_{\chi}\delta E}}=\frac{1}{\upsilon} \sigma_{0Z}\frac{m^2_{\chi}}{ m^2_e}  \left (C_{\ell,j_1,m_1,j_2,m_2}\right )^2\sqrt{\upsilon^2-\frac{2\delta E}{m_{\chi}}}
\eeq

We must now fold it with the velocity distribution in the local frame, ignoring the motion of the Earth around the sun, i.e.
\beq
f(\upsilon,\upsilon_0,\xi)=\frac{1}{\upsilon_0^3\pi \sqrt{\pi}}e^{-\frac{\upsilon^2+2\upsilon\upsilon_0\xi+\upsilon_0^2}{\upsilon_0^2}}
\label{Eq:fyxi}
\eeq
The integral over $\xi$ is done analytically to yield:
\beq
\langle (\sigma y)\rangle =\sigma_{0Z}  (C_{\ell,j_1,m_1,j_2,m_2})^2\frac{m_{\chi }^2}{ m^2_e}\frac{4}{\sqrt{\pi}}\int_{b}^{y_{max}} dy y y^2  e^{-y^2-1} \frac{\sinh (2y)}{2 y}
	\sqrt{1-\frac{b^2}{y^2}},\, b=\sqrt{\frac{2 \delta E}{m_{\chi}v_0^2}}
\eeq
or
\barr
\langle (\sigma y)\rangle &=&\sigma_{0Z}  (C_{\ell,j_1,m_1,j_2,m_2})^2\nonumber\\& &x^2 \frac{2}{\sqrt{\pi}}\int_{b}^{y_{max}} dy  y^2  e^{-y^2-1} \sinh (2y)
	\sqrt{1-\frac{b^2}{y^2}},\, b=\sqrt{\frac{2 \delta E}{x m_{e}v_0^2}},\, x=\frac{m_{\chi}}{m_e}
	\earr
The last integral can only be done numerically.
	
	The event rate, omitting the orbit dependent angular momentum coefficient $(C_{\ell,j_1,m_1,j_2,m_2})^2$takes the form:
	\beq
	R=\frac{\Lambda}{x}   x^2\frac{2}{\sqrt{\pi}} \int_{b}^{y_{max}} dy  y^2  e^{-y^2-1} \sinh (2y)
	\sqrt{1-\frac{b^2}{y^2}},\,b=\sqrt{\frac{7.3\left(\delta E/1 \mbox{eV}\right )}{x}}
	\label{Eq:atomrate5}
	\eeq
	where $\Lambda$ is  defined as 
	\beq
	\Lambda= \frac{\rho_{\chi}}{m_{e}} \sigma_0 {\upsilon_0} N_e 
	\label{Eq:elrate3b}
	\eeq
		One can easily find that the  constraint among the parameters is 
		$$\sqrt{\frac{7.3\left(\delta E/1 \mbox{eV}\right )}{x}}<2.84\Rightarrow x>0.9\frac{ \delta E}{1 \mbox{eV}}$$
	
\begin{figure}[!ht]
  \begin{center}
\subfloat[]
{
\includegraphics[width=0.4\textwidth,height=0.3\textwidth]{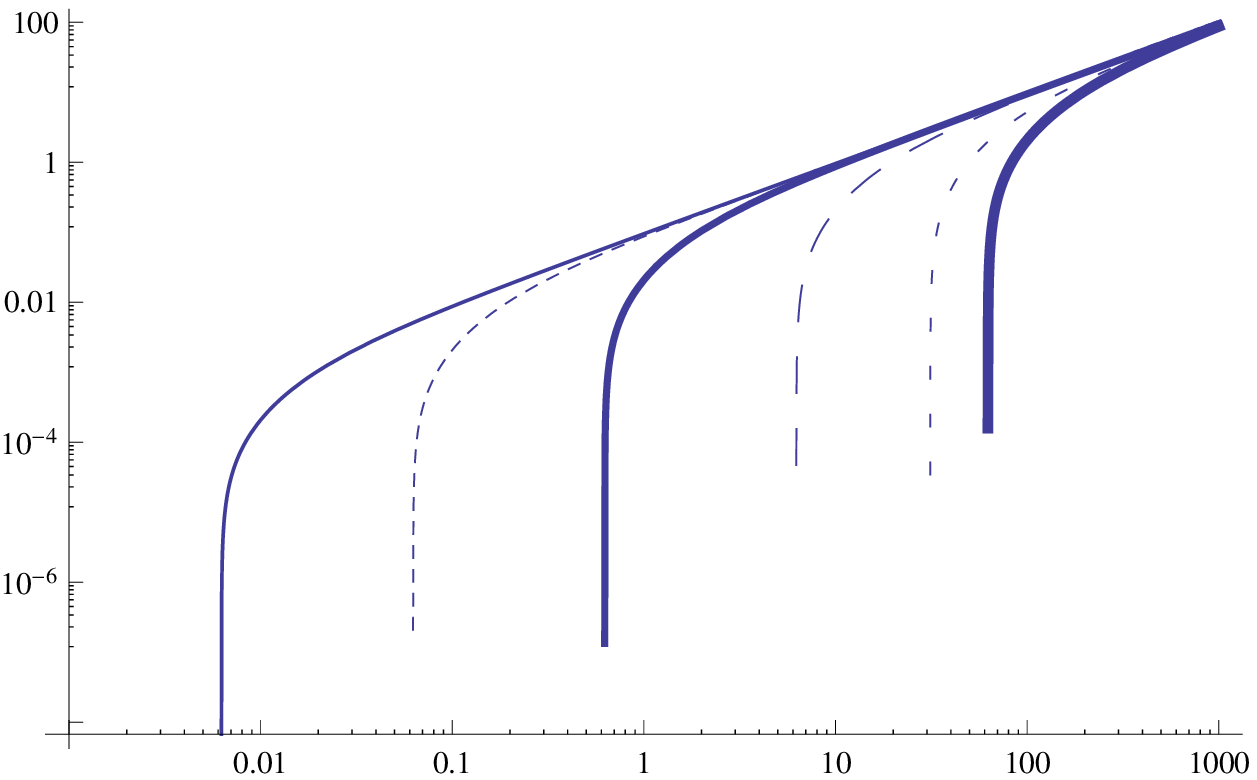}
\hspace{0.5cm}
}
\subfloat[]
{
\includegraphics[width=0.4\textwidth,height=0.3\textwidth]{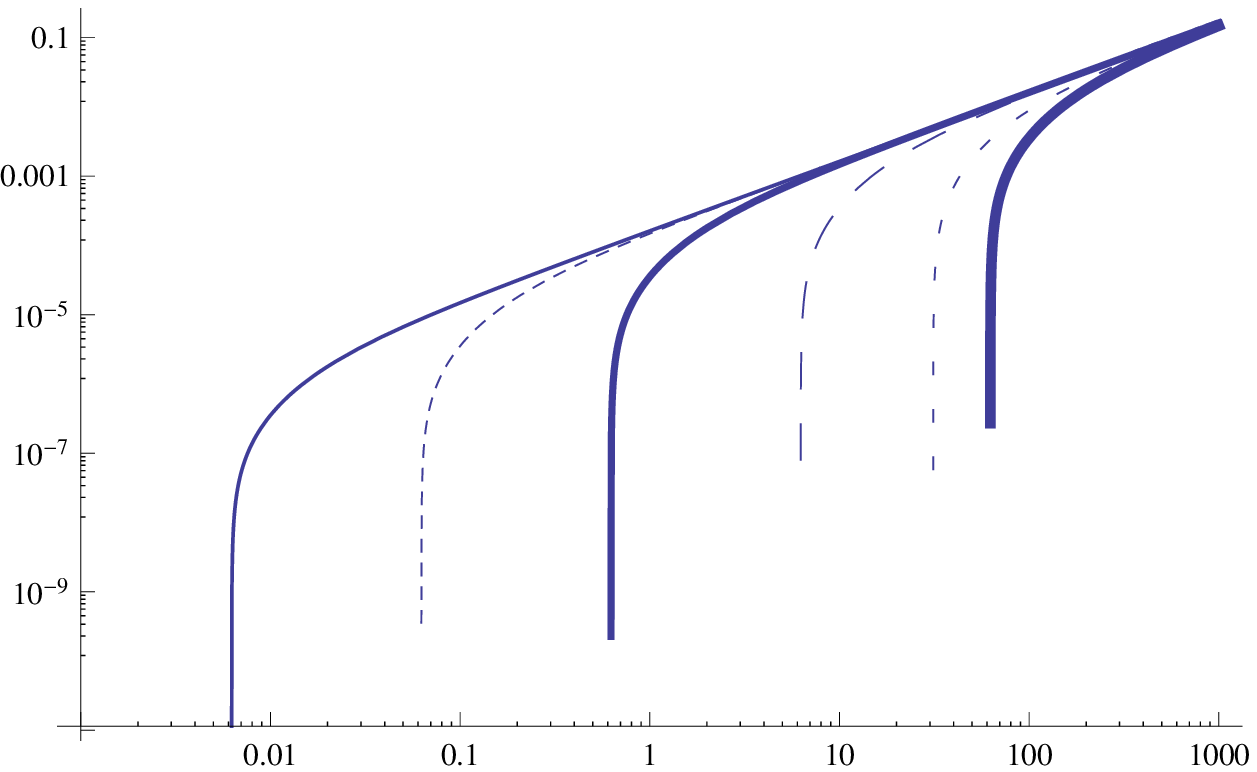}
}\\
\hspace*{-1.0cm}$\rightarrow x=\frac{m_{\chi}}{m_e}$	
 \caption{The total event rate  per year for a target with $N_A=10^{24}$ atoms as a function of  $x=\frac{m_x}{m_e}$ in the case of atomic excitations. In panel (a) the various curves correspond to   rates with $\Lambda=1$, while in panel (b) those with  the actual value of $\Lambda=1.7 \times 10^{-3}$.  In both cases the  curves correspond to values of $\delta$E= \{
   0.001,0.01,0.1,1.,5.,10.\} eV   increasing from left to right. 
 \label{fig:rateatom}}
 \end{center}
  \end{figure}

	The extra factor of $1/x$ in Eq. (\ref{Eq:atomrate5}) comes from the fact that the value of  $\Lambda$ employed has been evaluated with WIMP number density associated with a mass $m_e$, rather than $m_{\chi}$. It has, of course, been  assumed  one electron per atom $N_e=N_a=1.0\times 10^{24}$. We exhibit the obtained rates in Fig. \ref{fig:rateatom}.
	
	It is worth comparing the results obtained above with those of in the  of WIMP-electron scattering, see Fig. \ref{fig:totrate}. We see that for a given excitation energy the atomic rates increase with the WIMP mass. Thus, e.g.,  for $m_{\chi}=m_e$ the electron scattering yields $0.36\times 1.7\times 10^{-3}=6.0 \times 10^{-4}$ events per year. We will compare this with that associated  with   $\delta$E=0.1 eV excitation. We get $3.5 \times 10^{-5}$, 0.0015 and 0.014 for $x=1,10,100$ respectively. In other words the ratio of  atomic to recoil  events per year for free electrons becomes $0.06$,25 and $2.3\times^{3}$ in the above order for $x$.  Clearly the atomic excitations are much favored for  $x>5$. They are also much favored compared to detecting the recoi of bound electrons for light WIMPs. 
An additional advantage of the atomic experiments is the fact that targets with  a number of electrons $N_e>10^{24}$ are feasible.

The detection involves measuring the de-excitation of the populated level. It is also possible, following Sikivie's ideas \cite{Sikivie14} for axion detection,  to concentrate \cite{VerAvCres18} on the population of a preferred atomic level at low excitation provided that it is not otherwise occupied by electrons. Then shine a tunable laser to further excite the electrons to a preferred level and then obseve the de-excitation of the chosen level. This may require to cool system at very low temperatures and use a target, perhaps  enriched with an impurity if necessary, so that the system  maintains an atomic structure at the necessary  low temperature.
	
	
	The obtained rates in Fig. \ref{fig:rateatom} are in principle detectable, but it should be noted that the angular momentum factors $(C_{\ell,j_1,m_1,j_2,m_2})^2$ have not been included. These are tabulated in \ref{tab:tab1}-\ref{tab:tab2}.
\begin{table}
\caption{the coefficients $\left (C_{j_1,m_1,j_2,m_2,\ell}\right)^2 $ connecting via the spin operator a given initial state
 $|i\rangle=|n\ell,j_1,m_1\rangle$ with all possible states $|f \rangle=|n\ell,j_2,m_2\rangle $, for $\ell=0,\,1$. Note s-states are favored.}
\label{tab:tab1}
$$
\left(
\begin{array}{ccccc|c}
\ell&j_1&m_1&j_2&m_2&C^2_{j_1,m_1,j_2,m_2,\ell}\\
\hline
 0 & \frac{1}{2} & -\frac{1}{2} & \frac{1}{2} & \frac{1}{2} & 2 \\
\end{array}
\right),
\left (
\begin{array}{ccccc|c}
&|i\rangle&&|f\rangle&&\\
\hline
\ell&j_1&m_1&j_2&m_2&C^2_{j_1,m_1,j_2,m_2,\ell}\\
\hline
 1 & \frac{1}{2} & -\frac{1}{2} & \frac{1}{2} & \frac{1}{2} &
   \frac{2}{9} \\
 1 & \frac{1}{2} & -\frac{1}{2} & \frac{3}{2} & -\frac{3}{2} &
   \frac{4}{3} \\
 1 & \frac{1}{2} & -\frac{1}{2} & \frac{3}{2} & -\frac{1}{2} &
   \frac{8}{9} \\
 1 & \frac{1}{2} & -\frac{1}{2} & \frac{3}{2} & \frac{1}{2} &
   \frac{4}{9} \\
 1 & \frac{1}{2} & \frac{1}{2} & \frac{3}{2} & -\frac{1}{2} &
   \frac{4}{9} \\
 1 & \frac{1}{2} & \frac{1}{2} & \frac{3}{2} & \frac{1}{2} &
   \frac{8}{9} \\
 1 & \frac{1}{2} & \frac{1}{2} & \frac{3}{2} & \frac{3}{2} & \frac{4}{3}
   \\
 1 & \frac{3}{2} & -\frac{3}{2} & \frac{3}{2} & -\frac{1}{2} &
   \frac{2}{3} \\
 1 & \frac{3}{2} & -\frac{1}{2} & \frac{3}{2} & \frac{1}{2} &
   \frac{8}{9} \\
 1 & \frac{3}{2} & \frac{1}{2} & \frac{3}{2} & \frac{3}{2} & \frac{2}{3}
   \\
\end{array}
\right)
$$
\end{table}
\begin{table}
\caption{The same as in table \ref{tab:tab1}, the coefficients $\left (C_{j_1,m_1,j_2,m_2,\ell}\right )^2$ for $\ell=2$}
\label{tab:tab2}
$$
\left(
\begin{array}{ccccc|c}
&|i\rangle&&|f\rangle&&\\
\hline
\ell&j_1&m_1&j_2&m_2&C^2_{j_1,m_1,j_2,m_2,\ell}\\
\hline
 2 & \frac{3}{2} & -\frac{3}{2} & \frac{3}{2} & -\frac{1}{2} &
   \frac{6}{25} \\
 2 & \frac{3}{2} & -\frac{3}{2} & \frac{5}{2} & -\frac{5}{2} &
   \frac{8}{5} \\
 2 & \frac{3}{2} & -\frac{3}{2} & \frac{5}{2} & -\frac{3}{2} &
   \frac{16}{25} \\
 2 & \frac{3}{2} & -\frac{3}{2} & \frac{5}{2} & -\frac{1}{2} &
   \frac{4}{25} \\
 2 & \frac{3}{2} & -\frac{1}{2} & \frac{3}{2} & \frac{1}{2} &
   \frac{8}{25} \\
 2 & \frac{3}{2} & -\frac{1}{2} & \frac{5}{2} & -\frac{3}{2} &
   \frac{24}{25} \\
 2 & \frac{3}{2} & -\frac{1}{2} & \frac{5}{2} & -\frac{1}{2} &
   \frac{24}{25} \\
 2 & \frac{3}{2} & -\frac{1}{2} & \frac{5}{2} & \frac{1}{2} &
   \frac{12}{25} \\
 2 & \frac{3}{2} & \frac{1}{2} & \frac{3}{2} & \frac{3}{2} &
   \frac{6}{25} \\
 2 & \frac{3}{2} & \frac{1}{2} & \frac{5}{2} & -\frac{1}{2} &
   \frac{12}{25} \\
 2 & \frac{3}{2} & \frac{1}{2} & \frac{5}{2} & \frac{1}{2} &
   \frac{24}{25} \\
 2 & \frac{3}{2} & \frac{1}{2} & \frac{5}{2} & \frac{3}{2} &
   \frac{24}{25} \\
 2 & \frac{3}{2} & \frac{3}{2} & \frac{5}{2} & \frac{1}{2} &
   \frac{4}{25} \\
 2 & \frac{3}{2} & \frac{3}{2} & \frac{5}{2} & \frac{3}{2} &
   \frac{16}{25} \\
 2 & \frac{3}{2} & \frac{3}{2} & \frac{5}{2} & \frac{5}{2} &
   \frac{8}{5} \\
 2 & \frac{5}{2} & -\frac{5}{2} & \frac{5}{2} & -\frac{3}{2} &
   \frac{2}{5} \\
 2 & \frac{5}{2} & -\frac{3}{2} & \frac{5}{2} & -\frac{1}{2} &
   \frac{16}{25} \\
 2 & \frac{5}{2} & -\frac{1}{2} & \frac{5}{2} & \frac{1}{2} &
   \frac{18}{25} \\
 2 & \frac{5}{2} & \frac{1}{2} & \frac{5}{2} & \frac{3}{2} &
   \frac{16}{25} \\
 2 & \frac{5}{2} & \frac{3}{2} & \frac{5}{2} & \frac{5}{2} & \frac{2}{5}
   \\
\end{array}
\right)
 $$
\end{table}
	\section{Discussion}
	\label{sec:discussion}
	In the present paper we examined the possibility of detecting light WIMPs by exploiting their possible interactions with electrons. We found that, for WIMPs in the mass range of the electron mass, the electron recoiling energies are in the eV region. It is therefore very difficult for electrons  to be ejected by overcoming  their binding. Furthermore, for WIMP masses less than 50 times the electron mass,  the expected rate is too small to be observed.  Scattered electrons may be observed, if they are essentially free, with the use of electron detectors may be a good way to directly detect light WIMPs in the sub-MeV region. The WIMP density in our vicinity becomes quite high due to their small mass and  the  WIMP-electron cross section section may be quite enhanced for scalar WIMPs.
		
Such detectors  utilizing Fermi-degenerate materials like superconductors\cite{HPZ15} have recently been suggested. In this case the energy required is essentially the gap energy of about $1.5 kT_c$ which is in the meV region, i.e the electrons are essentially free. We have seen that event rates can be quite high for very light WIMPs, but the amount of energy deposited in the detector  is quite small. 

We have also seen that it may be possible to detect light WIMPs using a detector in a magnetic field via atomic excitations due to the well known electron  spin interactions

\section*{Acknowledgments} 
J.D.V  is happy to acknowledge support of this work by 
  the National Experts Council of China via  a "Foreign Master" grant.

\section*{References}

\begin{thebibliography}{61}
\expandafter\ifx\csname natexlab\endcsname\relax\def\natexlab#1{#1}\fi
\expandafter\ifx\csname bibnamefont\endcsname\relax
  \def\bibnamefont#1{#1}\fi
\expandafter\ifx\csname bibfnamefont\endcsname\relax
  \def\bibfnamefont#1{#1}\fi
\expandafter\ifx\csname citenamefont\endcsname\relax
  \def\citenamefont#1{#1}\fi
\expandafter\ifx\csname url\endcsname\relax
  \def\url#1{\texttt{#1}}\fi
\expandafter\ifx\csname urlprefix\endcsname\relax\def\urlprefix{URL }\fi
\providecommand{\bibinfo}[2]{#2}
\providecommand{\eprint}[2][]{\url{#2}}

\bibitem[{\citenamefont{Hanary et~al.}(2000)}]{MAXIMA1}
\bibinfo{author}{\bibfnamefont{S.}~\bibnamefont{Hanary}} \bibnamefont{et~al.},
  \bibinfo{journal}{Astrophys. J.} \textbf{\bibinfo{volume}{545}},
  \bibinfo{pages}{L5} (\bibinfo{year}{2000}).

\bibitem[{\citenamefont{Wu et~al.}(2001)}]{MAXIMA2}
\bibinfo{author}{\bibfnamefont{J.}~\bibnamefont{Wu}} \bibnamefont{et~al.},
  \bibinfo{journal}{Phys. Rev. Lett.} \textbf{\bibinfo{volume}{87}},
  \bibinfo{pages}{251303} (\bibinfo{year}{2001}).

\bibitem[{\citenamefont{Santos et~al.}(2002)}]{MAXIMA3}
\bibinfo{author}{\bibfnamefont{M.}~\bibnamefont{Santos}} \bibnamefont{et~al.},
  \bibinfo{journal}{Phys. Rev. Lett.} \textbf{\bibinfo{volume}{88}},
  \bibinfo{pages}{241302} (\bibinfo{year}{2002}).

\bibitem[{\citenamefont{Mauskopf et~al.}(2002)}]{BOOMERANG1}
\bibinfo{author}{\bibfnamefont{P.~D.} \bibnamefont{Mauskopf}}
  \bibnamefont{et~al.}, \bibinfo{journal}{Astrophys. J.}
  \textbf{\bibinfo{volume}{536}}, \bibinfo{pages}{L59} (\bibinfo{year}{2002}).

\bibitem[{\citenamefont{Mosi et~al.}(2002)}]{BOOMERANG2}
\bibinfo{author}{\bibfnamefont{S.}~\bibnamefont{Mosi}} \bibnamefont{et~al.},
  \bibinfo{journal}{Prog. Nuc.Part. Phys.} \textbf{\bibinfo{volume}{48}},
  \bibinfo{pages}{243} (\bibinfo{year}{2002}).

\bibitem[{\citenamefont{Halverson et~al.}(2002)}]{DASI02}
\bibinfo{author}{\bibfnamefont{N.~W.} \bibnamefont{Halverson}}
  \bibnamefont{et~al.}, \bibinfo{journal}{Astrophys. J.}
  \textbf{\bibinfo{volume}{568}}, \bibinfo{pages}{38} (\bibinfo{year}{2002}).

\bibitem[{\citenamefont{Smoot et~al.}(1992)}]{COBE}
\bibinfo{author}{\bibfnamefont{G.~F.} \bibnamefont{Smoot}}
  \bibnamefont{et~al.}, \bibinfo{journal}{Astrophys. J.}
  \textbf{\bibinfo{volume}{396}}, \bibinfo{pages}{L1} (\bibinfo{year}{1992}),
  \bibinfo{note}{the COBE Collaboration}.

\bibitem[{\citenamefont{Jaffe et~al.}(2001)}]{flat01}
\bibinfo{author}{\bibfnamefont{A.~H.} \bibnamefont{Jaffe}}
  \bibnamefont{et~al.}, \bibinfo{journal}{Phys. Rev. Lett.}
  \textbf{\bibinfo{volume}{86}}, \bibinfo{pages}{3475} (\bibinfo{year}{2001}).

\bibitem[{\citenamefont{Spergel et~al.}(2003)}]{SPERGEL}
\bibinfo{author}{\bibfnamefont{D.~N.} \bibnamefont{Spergel}}
  \bibnamefont{et~al.}, \bibinfo{journal}{Astrophys. J. Suppl.}
  \textbf{\bibinfo{volume}{148}}, \bibinfo{pages}{175} (\bibinfo{year}{2003}).

\bibitem[{\citenamefont{Spergel et~al.}(2007)}]{WMAP06}
\bibinfo{author}{\bibfnamefont{D.}~\bibnamefont{Spergel}} \bibnamefont{et~al.},
  \bibinfo{journal}{Astrophys. J. Suppl.} \textbf{\bibinfo{volume}{170}},
  \bibinfo{pages}{377} (\bibinfo{year}{2007}),
  \bibinfo{note}{[arXiv:astro-ph/0603449v2]}.

\bibitem[{Pla()}]{PlanckCP13}
\bibinfo{note}{The Planck Collaboration, A.P.R. Ade {\it et al}, Astron.
  Astrophys. 571 (2014), A16 arXiv:1303.5076 [astro-ph.CO]}.

\bibitem[{\citenamefont{Bennett et~al.}(1995)}]{Benne}
\bibinfo{author}{\bibfnamefont{D.~P.} \bibnamefont{Bennett}}
  \bibnamefont{et~al.}, \bibinfo{journal}{Phys. Rev. Lett.}
  \textbf{\bibinfo{volume}{74}}, \bibinfo{pages}{2867} (\bibinfo{year}{1995}).

\bibitem[{\citenamefont{Ullio and Kamioknowski}(2001)}]{UK01}
\bibinfo{author}{\bibfnamefont{P.}~\bibnamefont{Ullio}} \bibnamefont{and}
  \bibinfo{author}{\bibfnamefont{M.}~\bibnamefont{Kamioknowski}},
  \bibinfo{journal}{JHEP} \textbf{\bibinfo{volume}{0103}}, \bibinfo{pages}{049}
  (\bibinfo{year}{2001}).

\bibitem[{\citenamefont{Lewin and Smith}(1996)}]{LS96}
\bibinfo{author}{\bibfnamefont{J.~D.} \bibnamefont{Lewin}} \bibnamefont{and}
  \bibinfo{author}{\bibfnamefont{P.~F.} \bibnamefont{Smith}},
  \bibinfo{journal}{Astropart. Phys.} \textbf{\bibinfo{volume}{6}},
  \bibinfo{pages}{87} (\bibinfo{year}{1996}).

\bibitem[{\citenamefont{Goodman and Witten}(1985)}]{GOODWIT}
\bibinfo{author}{\bibfnamefont{M.~W.} \bibnamefont{Goodman}} \bibnamefont{and}
  \bibinfo{author}{\bibfnamefont{E.}~\bibnamefont{Witten}},
  \bibinfo{journal}{Phys. Rev. D} \textbf{\bibinfo{volume}{31}},
  \bibinfo{pages}{3059} (\bibinfo{year}{1985}).

\bibitem[{\citenamefont{Drukier et~al.}(1986)\citenamefont{Drukier, Freeze, and
  Spergel}}]{Druck}
\bibinfo{author}{\bibfnamefont{A.}~\bibnamefont{Drukier}},
  \bibinfo{author}{\bibfnamefont{K.}~\bibnamefont{Freeze}}, \bibnamefont{and}
  \bibinfo{author}{\bibfnamefont{D.}~\bibnamefont{Spergel}},
  \bibinfo{journal}{Phys. Rev. D} \textbf{\bibinfo{volume}{33}},
  \bibinfo{pages}{3495} (\bibinfo{year}{1986}).

\bibitem[{\citenamefont{Primack et~al.}(1988)\citenamefont{Primack, Seckel, and
  Sadoulet}}]{PSS88}
\bibinfo{author}{\bibfnamefont{J.~R.} \bibnamefont{Primack}},
  \bibinfo{author}{\bibfnamefont{D.}~\bibnamefont{Seckel}}, \bibnamefont{and}
  \bibinfo{author}{\bibfnamefont{B.}~\bibnamefont{Sadoulet}},
  \bibinfo{journal}{Ann. Rev. Nucl. Part. Sci.} \textbf{\bibinfo{volume}{38}},
  \bibinfo{pages}{751} (\bibinfo{year}{1988}).

\bibitem[{\citenamefont{Gabutti and Schmiemann}(1993)}]{GS93}
\bibinfo{author}{\bibfnamefont{A.}~\bibnamefont{Gabutti}} \bibnamefont{and}
  \bibinfo{author}{\bibfnamefont{K.}~\bibnamefont{Schmiemann}},
  \bibinfo{journal}{Phys. Lett. B} \textbf{\bibinfo{volume}{308}},
  \bibinfo{pages}{411} (\bibinfo{year}{1993}).

\bibitem[{\citenamefont{Bernabei}(1995)}]{RBERNABEI95}
\bibinfo{author}{\bibfnamefont{R.}~\bibnamefont{Bernabei}},
  \bibinfo{journal}{Riv. Nouvo Cimento} \textbf{\bibinfo{volume}{18 (5)}},
  \bibinfo{pages}{1} (\bibinfo{year}{1995}).

\bibitem[{\citenamefont{Abriola et~al.}(1999)}]{ABRIOLA98}
\bibinfo{author}{\bibfnamefont{D.}~\bibnamefont{Abriola}} \bibnamefont{et~al.},
  \bibinfo{journal}{Astropart. Phys.} \textbf{\bibinfo{volume}{10}},
  \bibinfo{pages}{133} (\bibinfo{year}{1999}),
  \bibinfo{note}{arXiv:astro-ph/9809018}.

\bibitem[{\citenamefont{Hasenbalg}(1998)}]{HASENBALG98}
\bibinfo{author}{\bibfnamefont{F.}~\bibnamefont{Hasenbalg}},
  \bibinfo{journal}{Astropart. Phys.} \textbf{\bibinfo{volume}{9}},
  \bibinfo{pages}{339} (\bibinfo{year}{1998}),
  \bibinfo{note}{arXiv:astro-ph/9806198}.

\bibitem[{\citenamefont{Vergados}(2003)}]{JDV03}
\bibinfo{author}{\bibfnamefont{J.~D.} \bibnamefont{Vergados}},
  \bibinfo{journal}{Phys. Rev. D} \textbf{\bibinfo{volume}{67}},
  \bibinfo{pages}{103003} (\bibinfo{year}{2003}),
  \bibinfo{note}{hep-ph/0303231}.

\bibitem[{\citenamefont{Green}(2003)}]{GREEN04}
\bibinfo{author}{\bibfnamefont{A.}~\bibnamefont{Green}},
  \bibinfo{journal}{Phys. Rev. D} \textbf{\bibinfo{volume}{68}},
  \bibinfo{pages}{023004} (\bibinfo{year}{2003}), \bibinfo{note}{ibid: D ${\bf
  69}$ (2004) 109902; arXiv:astro-ph/0304446}.

\bibitem[{\citenamefont{Savage et~al.}(2006)\citenamefont{Savage, Freese, and
  Gondolo}}]{SFG06}
\bibinfo{author}{\bibfnamefont{C.}~\bibnamefont{Savage}},
  \bibinfo{author}{\bibfnamefont{K.}~\bibnamefont{Freese}}, \bibnamefont{and}
  \bibinfo{author}{\bibfnamefont{P.}~\bibnamefont{Gondolo}},
  \bibinfo{journal}{Phys. Rev. D} \textbf{\bibinfo{volume}{74}},
  \bibinfo{pages}{043531} (\bibinfo{year}{2006}),
  \bibinfo{note}{arXiv:astro-ph/0607121}.

\bibitem[{FKL()}]{FKLW11}
\bibinfo{note}{P. J. Fox, J. Kopp, M. Lisanti and N. Weiner, A CoGeNT
  Modulation Analysis, arXiv:1107.0717 (astro-ph.CO)}.

\bibitem[{\citenamefont{Abe et~al.}(2009)}]{XMASS09}
\bibinfo{author}{\bibfnamefont{K.}~\bibnamefont{Abe}} \bibnamefont{et~al.},
  \bibinfo{journal}{Astropart. Phys.} \textbf{\bibinfo{volume}{31}},
  \bibinfo{pages}{290} (\bibinfo{year}{2009}), \bibinfo{note}{arXiv:v3
  [physics.ins-det]0809.4413v3 [physics.ins-det]}.

\bibitem[{\citenamefont{Bruch et~al.}(2009)}]{CDMSII09}
\bibinfo{author}{\bibfnamefont{T.}~\bibnamefont{Bruch}} \bibnamefont{et~al.},
  \bibinfo{journal}{Phys. Rev. Lett} \textbf{\bibinfo{volume}{102}},
  \bibinfo{pages}{011301} (\bibinfo{year}{2009}),
  \bibinfo{note}{arXiv:0802.3530}.

\bibitem[{\citenamefont{Armengaud et~al.}(2011)}]{EDELWEISS11}
\bibinfo{author}{\bibfnamefont{E.}~\bibnamefont{Armengaud}}
  \bibnamefont{et~al.}, \bibinfo{journal}{Phys. Lett. B}
  \textbf{\bibinfo{volume}{702}}, \bibinfo{pages}{329} (\bibinfo{year}{2011}),
  \bibinfo{note}{arXiv:1103.4070v3 [astro-ph.CO]}.

\bibitem[{\citenamefont{Kim et~al.}(2012)}]{KIMS12}
\bibinfo{author}{\bibfnamefont{S.~C.} \bibnamefont{Kim}} \bibnamefont{et~al.},
  \bibinfo{journal}{Phys. Rev. Lett.} \textbf{\bibinfo{volume}{108}},
  \bibinfo{pages}{181301} (\bibinfo{year}{2012}), \bibinfo{note}{for the KIMS
  collaboration}.

\bibitem[{\citenamefont{Felizardo et~al.}(2012)}]{SIMPLE12}
\bibinfo{author}{\bibfnamefont{M.}~\bibnamefont{Felizardo}}
  \bibnamefont{et~al.}, \bibinfo{journal}{Phys. Rev. Lett.}
  \textbf{\bibinfo{volume}{108}}, \bibinfo{pages}{201302}
  (\bibinfo{year}{2012}), \bibinfo{note}{the SIMPLE collaboration, see also
  Erratum: Phys. Rev. D 90, 079902 (2014), arXiv:1003.2987 [astro-phCO]}.

\bibitem[{\citenamefont{Archambault et~al.}(2012)}]{PICASSO12}
\bibinfo{author}{\bibfnamefont{S.}~\bibnamefont{Archambault}}
  \bibnamefont{et~al.}, \bibinfo{journal}{Phys. Lett. B}
  \textbf{\bibinfo{volume}{711}}, \bibinfo{pages}{153} (\bibinfo{year}{2012}),
  \bibinfo{note}{the PICASSO collaboration}.

\bibitem[{\citenamefont{Bernabei et~al.}(2013)}]{DAMAEPJ13}
\bibinfo{author}{\bibfnamefont{R.}~\bibnamefont{Bernabei}}
  \bibnamefont{et~al.}, \bibinfo{journal}{Eur. Phys. J.}
  \textbf{\bibinfo{volume}{C 73}}, \bibinfo{pages}{2648}
  (\bibinfo{year}{2013}), \bibinfo{note}{[DAMA/LIBRA phase 1]; arXiv:1308.5109
  (astro-GA.)}.

\bibitem[{CRE()}]{CRESST}
\bibinfo{note}{The CRESST Experiment: Recent Results and Prospects, P.Di
  Stefano, {\it et al}, arXiv:hep-ex/0011064; The CRESST Collaboration, talk
  presented at IBS - MultiDark Joint Focus Program, Daejeon, s. Korea, 10 –
  21 October 2014}.

\bibitem[{\citenamefont{Aprile et~al.}(2017)}]{XENON10017}
\bibinfo{author}{\bibfnamefont{E.}~\bibnamefont{Aprile}} \bibnamefont{et~al.},
  \bibinfo{journal}{Phys. Rev. Lett.} \textbf{\bibinfo{volume}{119}},
  \bibinfo{pages}{181301} (\bibinfo{year}{2017}), \bibinfo{note}{[XENON100
  Collaboration]; [arXiv:1705.06655]}.

\bibitem[{\citenamefont{Akerib et~al.}(2014)}]{LUX14}
\bibinfo{author}{\bibfnamefont{D.}~\bibnamefont{Akerib}} \bibnamefont{et~al.},
  \bibinfo{journal}{Phys. Rev. Lett.} \textbf{\bibinfo{volume}{112}},
  \bibinfo{pages}{091303} (\bibinfo{year}{2014}),
  \bibinfo{note}{[arXiv:1310.8214]}.

\bibitem[{KST()}]{KSTH18}
\bibinfo{note}{Sunghyun Kang, Stefano Scopel, Gaurav Tomar, Jong–Hyun,
  Present and projected sensitivities of Dark Matter direct detection
  experiments to effective WIMP-nucleus couplings, arXiv:1805.06113}.

\bibitem[{LUX()}]{LUXZEP}
\bibinfo{note}{LUX-ZEPLIN Collaboration, D. S. Akerib et al., Projected WIMP
  sensitivity of the LUX-ZEPLIN (LZ) dark matter experiment, arXiv:1802.06039}.

\bibitem[{\citenamefont{Essig et~al.}(2012{\natexlab{a}})\citenamefont{Essig,
  Mardon, and Volansky}}]{EMV12}
\bibinfo{author}{\bibfnamefont{R.}~\bibnamefont{Essig}},
  \bibinfo{author}{\bibfnamefont{J.}~\bibnamefont{Mardon}}, \bibnamefont{and}
  \bibinfo{author}{\bibfnamefont{T.}~\bibnamefont{Volansky}},
  \bibinfo{journal}{Phys. Rev. D} \textbf{\bibinfo{volume}{85}},
  \bibinfo{pages}{076007} (\bibinfo{year}{2012}{\natexlab{a}}).

\bibitem[{\citenamefont{Essig et~al.}(2012{\natexlab{b}})\citenamefont{Essig,
  Manalaysay, Mardon, and Volansky}}]{EMMPV12}
\bibinfo{author}{\bibfnamefont{R.}~\bibnamefont{Essig}},
  \bibinfo{author}{\bibfnamefont{A.}~\bibnamefont{Manalaysay}},
  \bibinfo{author}{\bibfnamefont{J.}~\bibnamefont{Mardon}}, \bibnamefont{and}
  \bibinfo{author}{\bibfnamefont{T.}~\bibnamefont{Volansky}},
  \bibinfo{journal}{Phys. Rev. Lett} \textbf{\bibinfo{volume}{109}},
  \bibinfo{pages}{021301} (\bibinfo{year}{2012}{\natexlab{b}}).

\bibitem[{\citenamefont{Schutz and Zurec}(2016)}]{SchZur16}
\bibinfo{author}{\bibfnamefont{K.}~\bibnamefont{Schutz}} \bibnamefont{and}
  \bibinfo{author}{\bibfnamefont{K.~M.} \bibnamefont{Zurec}},
  \bibinfo{journal}{Phys. Rev. Lett} \textbf{\bibinfo{volume}{117}},
  \bibinfo{pages}{121302} (\bibinfo{year}{2016}),
  \bibinfo{note}{arXiv:1604.0820}.

\bibitem[{\citenamefont{Hochberg et~al.}(2017)\citenamefont{Hochberg, Kahn,
  Lisanti, Tully, and Zurek}}]{HKLTZ17}
\bibinfo{author}{\bibfnamefont{Y.}~\bibnamefont{Hochberg}},
  \bibinfo{author}{\bibfnamefont{Y.}~\bibnamefont{Kahn}},
  \bibinfo{author}{\bibfnamefont{M.}~\bibnamefont{Lisanti}},
  \bibinfo{author}{\bibfnamefont{C.~G.} \bibnamefont{Tully}}, \bibnamefont{and}
  \bibinfo{author}{\bibfnamefont{K.~M.} \bibnamefont{Zurek}},
  \bibinfo{journal}{Phys. Lett. B} \textbf{\bibinfo{volume}{772}},
  \bibinfo{pages}{239} (\bibinfo{year}{2017}).

\bibitem[{\citenamefont{Derenzo et~al.}(2017)\citenamefont{Derenzo, Essig,
  Massari, Soto, and Yu}}]{DEMSY17}
\bibinfo{author}{\bibfnamefont{S.}~\bibnamefont{Derenzo}},
  \bibinfo{author}{\bibfnamefont{R.}~\bibnamefont{Essig}},
  \bibinfo{author}{\bibfnamefont{A.}~\bibnamefont{Massari}},
  \bibinfo{author}{\bibfnamefont{A.}~\bibnamefont{Soto}}, \bibnamefont{and}
  \bibinfo{author}{\bibfnamefont{T.-T.} \bibnamefont{Yu}},
  \bibinfo{journal}{Phys. Rev. D} \textbf{\bibinfo{volume}{96}},
  \bibinfo{pages}{016026} (\bibinfo{year}{2017}),
  \bibinfo{note}{arXiv:1703.01009 (astro-ph.CO), (hep-ex)}.

\bibitem[{Kop()}]{KopVer19}
\bibinfo{note}{G. Kopidakis, J.D. Vergados {\it et al}, to be published}.

\bibitem[{\citenamefont{Oikonomou et~al.}(2007)\citenamefont{Oikonomou,
  Vergados, and Moustakidis}}]{OikVerMou}
\bibinfo{author}{\bibfnamefont{V.}~\bibnamefont{Oikonomou}},
  \bibinfo{author}{\bibfnamefont{J.}~\bibnamefont{Vergados}}, \bibnamefont{and}
  \bibinfo{author}{\bibfnamefont{C.~C.} \bibnamefont{Moustakidis}},
  \bibinfo{journal}{Nuc. Phys.} \textbf{\bibinfo{volume}{B 773}},
  \bibinfo{pages}{19} (\bibinfo{year}{2007}).

\bibitem[{\citenamefont{Boehm and Fayet}(2004)}]{Fayet03}
\bibinfo{author}{\bibfnamefont{C.}~\bibnamefont{Boehm}} \bibnamefont{and}
  \bibinfo{author}{\bibfnamefont{P.}~\bibnamefont{Fayet}},
  \bibinfo{journal}{Nucl.Phys. B} \textbf{\bibinfo{volume}{683}},
  \bibinfo{pages}{29} (\bibinfo{year}{2004}),
  \bibinfo{note}{arXiv:hep-ph/0305261}.

\bibitem[{\citenamefont{Ma}(2006)}]{Ma06}
\bibinfo{author}{\bibfnamefont{E.}~\bibnamefont{Ma}}, \bibinfo{journal}{Phys.
  Rev. D} \textbf{\bibinfo{volume}{73}}, \bibinfo{pages}{077301}
  (\bibinfo{year}{2006}), \bibinfo{note}{arXiv:hep-ph/0601225}.

\bibitem[{\citenamefont{Silveira and Zee}(1985)}]{ZeeScal85}
\bibinfo{author}{\bibfnamefont{V.}~\bibnamefont{Silveira}} \bibnamefont{and}
  \bibinfo{author}{\bibfnamefont{A.}~\bibnamefont{Zee}},
  \bibinfo{journal}{Phys. Lett. B} \textbf{\bibinfo{volume}{161}},
  \bibinfo{pages}{136} (\bibinfo{year}{1985}).

\bibitem[{\citenamefont{Holz and Zee}(201)}]{ZeeScal01}
\bibinfo{author}{\bibfnamefont{D.}~\bibnamefont{Holz}} \bibnamefont{and}
  \bibinfo{author}{\bibfnamefont{A.}~\bibnamefont{Zee}},
  \bibinfo{journal}{Phys. Lett. B} \textbf{\bibinfo{volume}{517}},
  \bibinfo{pages}{239} (\bibinfo{year}{201}).

\bibitem[{\citenamefont{Bento et~al.}(2001)\citenamefont{Bento, Berolami, and
  Rosefeld}}]{BentoRos01}
\bibinfo{author}{\bibfnamefont{M.}~\bibnamefont{Bento}},
  \bibinfo{author}{\bibfnamefont{O.}~\bibnamefont{Berolami}}, \bibnamefont{and}
  \bibinfo{author}{\bibfnamefont{R.}~\bibnamefont{Rosefeld}},
  \bibinfo{journal}{Phys. lett. B} \textbf{\bibinfo{volume}{518}},
  \bibinfo{pages}{276} (\bibinfo{year}{2001}).

\bibitem[{\citenamefont{Bento et~al.}(2000)\citenamefont{Bento, Berolami,
  Rosefeld, and Teodoro}}]{BentoBero00}
\bibinfo{author}{\bibfnamefont{M.}~\bibnamefont{Bento}},
  \bibinfo{author}{\bibfnamefont{O.}~\bibnamefont{Berolami}},
  \bibinfo{author}{\bibfnamefont{R.}~\bibnamefont{Rosefeld}}, \bibnamefont{and}
  \bibinfo{author}{\bibfnamefont{L.}~\bibnamefont{Teodoro}},
  \bibinfo{journal}{Phys. Rev. D} \textbf{\bibinfo{volume}{62}},
  \bibinfo{pages}{041302} (\bibinfo{year}{2000}).

\bibitem[{\citenamefont{Cheung and Vergados}(2015)}]{Cheung:2014pea}
\bibinfo{author}{\bibfnamefont{Y.-K.~E.} \bibnamefont{Cheung}}
  \bibnamefont{and} \bibinfo{author}{\bibfnamefont{J.~D.}
  \bibnamefont{Vergados}}, \bibinfo{journal}{JCAP}
  \textbf{\bibinfo{volume}{1502}}, \bibinfo{pages}{014} (\bibinfo{year}{2015}),
  \eprint{1410.5710}.

\bibitem[{\citenamefont{Li et~al.}(2014)\citenamefont{Li, Brandenberger, and
  Cheung}}]{Li:2014era}
\bibinfo{author}{\bibfnamefont{C.}~\bibnamefont{Li}},
  \bibinfo{author}{\bibfnamefont{R.~H.} \bibnamefont{Brandenberger}},
  \bibnamefont{and} \bibinfo{author}{\bibfnamefont{Y.-K.~E.}
  \bibnamefont{Cheung}}, \bibinfo{journal}{Phys. Rev.}
  \textbf{\bibinfo{volume}{D90}}, \bibinfo{pages}{123535}
  (\bibinfo{year}{2014}), \eprint{1403.5625}.

\bibitem[{\citenamefont{Cheung et~al.}(2014)\citenamefont{Cheung, Kang, and
  Li}}]{Cheung:2014nxi}
\bibinfo{author}{\bibfnamefont{Y.-K.~E.} \bibnamefont{Cheung}},
  \bibinfo{author}{\bibfnamefont{J.~U.} \bibnamefont{Kang}}, \bibnamefont{and}
  \bibinfo{author}{\bibfnamefont{C.}~\bibnamefont{Li}}, \bibinfo{journal}{JCAP}
  \textbf{\bibinfo{volume}{1411}}, \bibinfo{pages}{001} (\bibinfo{year}{2014}),
  \eprint{1408.4387}.

\bibitem[{\citenamefont{Vergados et~al.}(2018)\citenamefont{Vergados,
  Moustakidis, Cheung, Ejiri, Kim, and yen Lee}}]{VMCEKL18}
\bibinfo{author}{\bibfnamefont{J.}~\bibnamefont{Vergados}},
  \bibinfo{author}{\bibfnamefont{C.~C.} \bibnamefont{Moustakidis}},
  \bibinfo{author}{\bibfnamefont{Y.-K.~E.} \bibnamefont{Cheung}},
  \bibinfo{author}{\bibfnamefont{H.}~\bibnamefont{Ejiri}},
  \bibinfo{author}{\bibfnamefont{Y.}~\bibnamefont{Kim}}, \bibnamefont{and}
  \bibinfo{author}{\bibfnamefont{J.}~\bibnamefont{yen Lee}},
  \bibinfo{journal}{AHEP} \textbf{\bibinfo{volume}{2018}},
  \bibinfo{pages}{6257198} (\bibinfo{year}{2018}).

\bibitem[{\citenamefont{Angloher et~al.}(2014)}]{CRESSTTUM40}
\bibinfo{author}{\bibfnamefont{G.}~\bibnamefont{Angloher}}
  \bibnamefont{et~al.}, \bibinfo{journal}{Eur. Phys. J C}
  \textbf{\bibinfo{volume}{74}}, \bibinfo{pages}{12} (\bibinfo{year}{2014}),
  \bibinfo{note}{astro-ph/1407.3146}.

\bibitem[{HPZ()}]{HPZ15}
\bibinfo{note}{Y. Hocberg,M. Pyle, Y. Zhao and M, Zurek, Detecting superlight
  Dark Matter with Fermi Degenerate Materials, arXiv:1512,04533 [hep-ph]}.

\bibitem[{\citenamefont{Larkins}(1977)}]{Larkins77}
\bibinfo{author}{\bibfnamefont{F.}~\bibnamefont{Larkins}},
  \bibinfo{journal}{At. Data and Nucl. Data Tables}
  \textbf{\bibinfo{volume}{20}}, \bibinfo{pages}{313} (\bibinfo{year}{1977}).

\bibitem[{Sev()}]{Sevier72}
\bibinfo{note}{K.D. Sevier, Low Energy Electron Spectrometry, Wiley-
  Interscience, New York (1972)}.

\bibitem[{\citenamefont{F.T and Freedman}(1978)}]{PorFreed78}
\bibinfo{author}{\bibfnamefont{P.}~\bibnamefont{F.T}} \bibnamefont{and}
  \bibinfo{author}{\bibfnamefont{M.}~\bibnamefont{Freedman}},
  \bibinfo{journal}{J. Phys. Chem. Ref. Data} \textbf{\bibinfo{volume}{7}},
  \bibinfo{pages}{1267} (\bibinfo{year}{1978}).

\bibitem[{\citenamefont{Sikivie}(2014)}]{Sikivie14}
\bibinfo{author}{\bibfnamefont{P.}~\bibnamefont{Sikivie}},
  \bibinfo{journal}{Phys. Rev. Lett.} \textbf{\bibinfo{volume}{113}},
  \bibinfo{pages}{201301} (\bibinfo{year}{2014}).

\bibitem[{Ver()}]{VerAvCres18}
\bibinfo{note}{F.T. Avignone III, R. J. Creswick and J. D. Vergados, Axion
  Detection via Atomic Excitations, arXiv:1801.02072 (hep-ph)}.

\end{thebibliography}

	\end{document}